\documentclass[3p]{elsarticle}
\pdfoutput=1
\usepackage{epsfig}
\usepackage{amsmath}
\usepackage{amssymb}
\usepackage{wasysym}
\usepackage{leftidx}

\def\beq{\begin{equation}}
\def\eeq{\end{equation}}
\def\ba{\begin{eqnarray}}
\def\ea{\end{eqnarray}}
\def\bal{\begin{align}}
\def\eal{\end{align}}
\def\bxi{\boldsymbol{\xi}}
\def\bnab{\boldsymbol{\nabla}}

\begin{document}

\title{Saturn Ring Seismology: Evidence for Stable Stratification in the Deep Interior of Saturn}

\author[jf,jf2]{Jim Fuller}
\ead{jfuller@caltech.edu}

\address[jf]{Kavli Institute for Theoretical Physics, Kohn Hall, University of California, Santa Barbara, CA 93106, USA}
\address[jf2]{TAPIR, Mailcode 350-17, California Institute of Technology, Pasadena, CA 91125, USA}

\label{firstpage}

\begin{abstract}

Seismology allows for direct observational constraints on the interior structures of stars and planets. Recent observations of Saturn's ring system have revealed the presence of density waves within the rings excited by oscillation modes within Saturn, allowing for precise measurements of a limited set of the planet's mode frequencies. We construct interior structure models of Saturn, compute the corresponding mode frequencies, and compare them with the observed mode frequencies. The fundamental mode frequencies of our models match the observed frequencies (of the largest amplitude waves) to an accuracy of $\sim 1 \%$, confirming that these waves are indeed excited by Saturn's f-modes. The presence of the lower amplitude waves (finely split in frequency from the f-modes) can only be reproduced in models containing gravity modes that propagate in a stably stratified region of the planet. The stable stratification must exist deep within the planet near the large density gradients between the core and envelope. Our models cannot easily reproduce the observed fine splitting of the $m=-3$ modes, suggesting that additional effects (e.g., significant latitudinal differential rotation) may be important. 

\end{abstract}

\maketitle

\section{Introduction}

The interior structures of planets other than the Earth are generally poorly constrained. Although theoretical studies abound, our understanding is hampered by the lack of direct observational constraints (see Guillot 2005, Fortney \& Nettelman 2009, and Guillot \& Gautier 2014 for reviews). With thousands of recently discovered exoplanets/exoplanet candidates, a basic understanding of the internal structures of giant planets is more important than ever.

Seismology offers the best hope for directly inferring interior structures of planets. Indeed, our understanding of the Earth's interior owes its existence primarily to seismic measurements. We advise the interested reader to consult Dahlen \& Tromp (1998), hereafter DT98, for a comprehensive description of the techniques of Earth seismology. Chaplin \& Miglio (2013) presents a review of recent developments in asteroseismology, while Lognonne \& Mosser (1993) and Stein \& Wysession (2003) discuss results in terrestrial seismology. Unfortunately, seismic measurements of other planets are much more difficult, and no unambiguous detections of oscillations in the outer Solar System planets exist (although there are tentative detections of pressure modes in Jupiter via radial velocity techniques, see Gaulme et al. 2011 and discussion in Section \ref{modeamp}). 

Saturn provides an amazing opportunity to indirectly detect global oscillation modes through their interaction with Saturn's rings. Marley (1991) (M91) and Marley \& Porco 1993 (MP93) predicted that some of Saturn's oscillation modes (in particular the prograde f-modes) could be detected through waves in the rings launched at Lindblad resonances with the gravitational forcing created by the modes (see also Pena 2010). This prediction was confirmed by Hedman \& Nicholson (2013) (HN13), who used {\it Cassini} data to measure the azimuthal pattern numbers $m$ and pattern frequencies $\Omega_p$ of several unidentified waves within Saturn's C Ring. They found that the frequencies and pattern numbers matched Marley \& Porco's predictions to within a few percent, and that the waves could not be explained by any other known process. 

The puzzling aspect of HN13's findings is that there are {\it multiple} waves of the same $m$ near the locations predicted by M91 and MP93, whereas only one wave was expected. The multiple waves, split by less than $10\%$ in frequency, appear to be generated by distinct oscillation modes within Saturn whose frequencies are split by the same fractions. The observed splitting is not simple rotational splitting (as this occurs between oscillation modes of different $m$) and suggests more complex physics is at play.

Fuller et al. (2014) (F14) investigated the effect of a solid core on the oscillation mode spectrum of Saturn. They found that if Saturn has a large solid core that is relatively unrigid (has a small shear modulus $\mu$), the shear oscillation modes of the core can exist near the same frequencies as the f-modes that generate some of the observed waves in the rings. Modes very close in frequency to the f-modes can degenerately mix with them (a process also known as avoided crossing), attaining large enough gravitational potential perturbations to generate waves in the rings. However, F14 found that degenerate mixing was rare, and that only finely tuned models could qualitatively reproduce the observed mode spectrum. The oscillations of rotating giant planets have also been examined in several other works (Vorontsov \& Zharkov 1981, Vorontsov 1981, Vorontsov 1984, M91, Wu 2005, Pena 2010, Le
Bihan \& Burrows 2012, Jackiewicz et al. 2012, Braviner \& Ogilvie 2014). None of these works have extensively examined the effect of stable stratification and the resulting planetary mode spectrum (although M91 does briefly consider the effect of stable stratification on the f-mode frequencies).

\begin{figure}
\begin{center}
\includegraphics[scale=0.45]{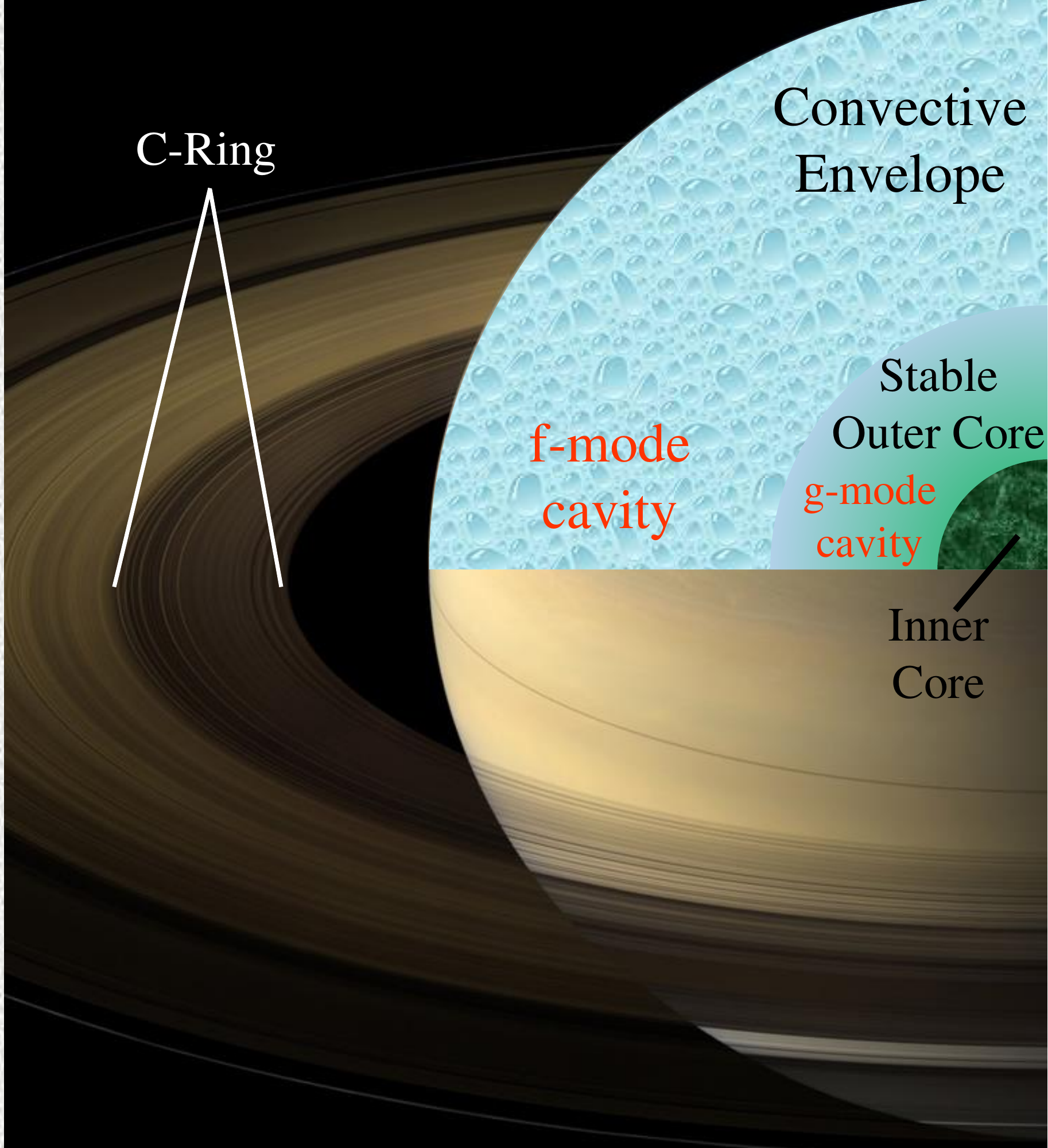}
\end{center} 
\caption{ \label{SatDiagram} {\it Cassini} image of Saturn and its rings, overlaid with a schematic cartoon of our hypothesis for Saturn's interior structure. The structure shown here is not quantitatively accurate. It is meant only to illustrate the general features of Saturn's interior structure that we advocate: a thick convective envelope (which harbors f-modes, p-modes, and i-modes) overlying a region of stable stratification  near the core of the planet (which harbors g-modes and r-modes). We have also pointed out the C-ring, where all of the mode-driven waves of been observed.}
\end{figure}

In this paper, we examine Saturn's oscillation mode spectrum in the presence of a stably stratified region deep within the planet. Regions of stable stratification have been speculated to exist within giant planets due to the stabilizing effect of composition gradients (Leconte \& Chabrier 2012, 2013). The composition gradients could be produced by dissolution of heavy core elements in the helium/hydrogren envelope (Wilson \& Militzer 2012a, 2012b) or by gravitational settling of metals (Stevenson 1985) or helium (Salpeter 1973, Stevenson \& Salpeter 1977). Recent simulations have sought to determine the large-scale time evolution of doubly diffusive convection produced by competing thermal/composition gradients (Rosenblum et al. 2011, Mirouh et al. 2012, Wood et al. 2013), but the resulting global structure of giant planets is unclear. Figure \ref{SatDiagram} shows a simple schematic of the type of Saturn models we consider. It should not be interpreted too strictly, it is intended only to provide the reader with a general picture of our hypothesis for Saturn's interior structure.

If stably stratified regions exist, they allow for the existence of gravity modes (g-modes) in the oscillation mode spectrum. For stable stratification deep within the planet, the g-modes can exist in the same frequency regime as the f-modes and can strongly mix with them. This process is analogous to the mixed g-modes/p-modes observed in red giant stars, although somewhat complicated by Saturn's rapid rotation. Our calculations reveal that g-mode mixing can naturally explain the observed splitting between the $m=-2$ waves, but cannot robustly reproduce the fine splitting between the $m=-3$ waves. We claim this is strong evidence for the existence of stable stratification within Saturn, although some important physical ingredient (e.g., differential rotation) may be required for a complete understanding.

Our paper is organized as follows. Section \ref{models} describes the toy Saturn models we use in our calculations. Section \ref{modes} summarizes our method of solving for oscillation modes in the presence of rapid rotation, and reviews the types of modes that exist in rotating planets. In Section \ref{mix} we examine the process of mode mixing induced by rotation, centrifugal, and ellipticity effects, and describe how this affects mode frequencies and eigenfunctions. Section \ref{results} compares our results to observations, and we conclude with a discussion of these results in Section \ref{discussion}. This section also addresses the issues of mode amplitudes, mode driving, and the prospects for observing Saturnian and Jovian p-modes via radial velocity techniques.

\section{Saturn Models}
\label{models}

The interior structure of giant planets is not particularly well-constrained. Other than its mass $M$ and radius $R$, the strongest observational constraint on Saturn's interior structure is the measured value of the gravitational moment $J_2$, which indicates that Saturn must have a dense core of $\sim 15 M_\oplus$ (Guillot 2005). We therefore create toy models which roughly match the measured values of $M$, equatorial radius $R_{\rm eq}$, polar radius $R_{\rm po}$, and $J_2$. We do not attempt to rigorously compare these models with any theoretical equations of state or microphysical models, although in Section \ref{discussion} we discuss how our models relate to recent theoretical developments in planetary interiors. 

For the purposes of our adiabatic mode calculations, the only physical quantities of importance in Saturn's interior are the density $\rho$, Brunt-Vaisala frequency $N$, sound speed $c_s$, gravitational acceleration $g$, and spin frequency $\Omega_s$. To create a toy model, we proceed as follows. We first create a spherical model with a polytropic density profile of index $(n=1)$, with a density profile $\rho_1(r)$. We choose a sound speed $c_{s1}$ such that the Brunt-Vaisala frequency
\beq
\label{brunt}
N_1^2 = -g_1 \frac{d {\rm ln} \rho_1}{d r} - \frac{g_1}{c_{s1}^2}
\eeq
is equal to zero everywhere. 

We then choose inner and outer core radii $r_{\rm in}$ and $r_{\rm out}$, and a core density enhancement $D$. We multiply the density of the inner core by $D$, such that $\rho(r) = D \rho_1(r)$ for $r<r_{\rm in}$. The density of the outer core is calculated via 
\beq
\rho(r) = \rho_1(r) \bigg[1 + (D-1) \sin^2\big[(\pi/2) (r_{\rm out}-r)/(r_{\rm out}-r_{\rm in})\big] \bigg] \qquad {\rm for} \quad r_{\rm in} < r < r_{\rm out}
\eeq
This form is somewhat arbitrary; we use it to obtain a smooth density increase between the envelope and inner core. In the outer core, we readjust the soundspeed such that
\beq
\label{cs}
c_s^2(r) = c_{s1}^2(r_{\rm out}) + \big[c_{s1}^2(r_{\rm in})-c_{s1}^2(r_{\rm out})\big] \sin \big[ (\pi/2) (r_{\rm out}-r)/(r_{\rm out}-r_{\rm in})\big] \qquad {\rm for} \quad r_{\rm in} < r < r_{\rm out}. 
\eeq
Once again, this sound speed profile is somewhat arbitrary. This form ensures a positive value of $N^2$ in the outer core. Because we focus only on f-modes (for which $\rho$ is the defining quantity) and g-modes (for which $N^2$ is the defining quantity) the precise value of $c_s^2$ has little effect on our results, except in so far as it affects the value of $N^2$ through equation \ref{brunt}. 

\begin{figure}
\begin{center}
\includegraphics[scale=0.45]{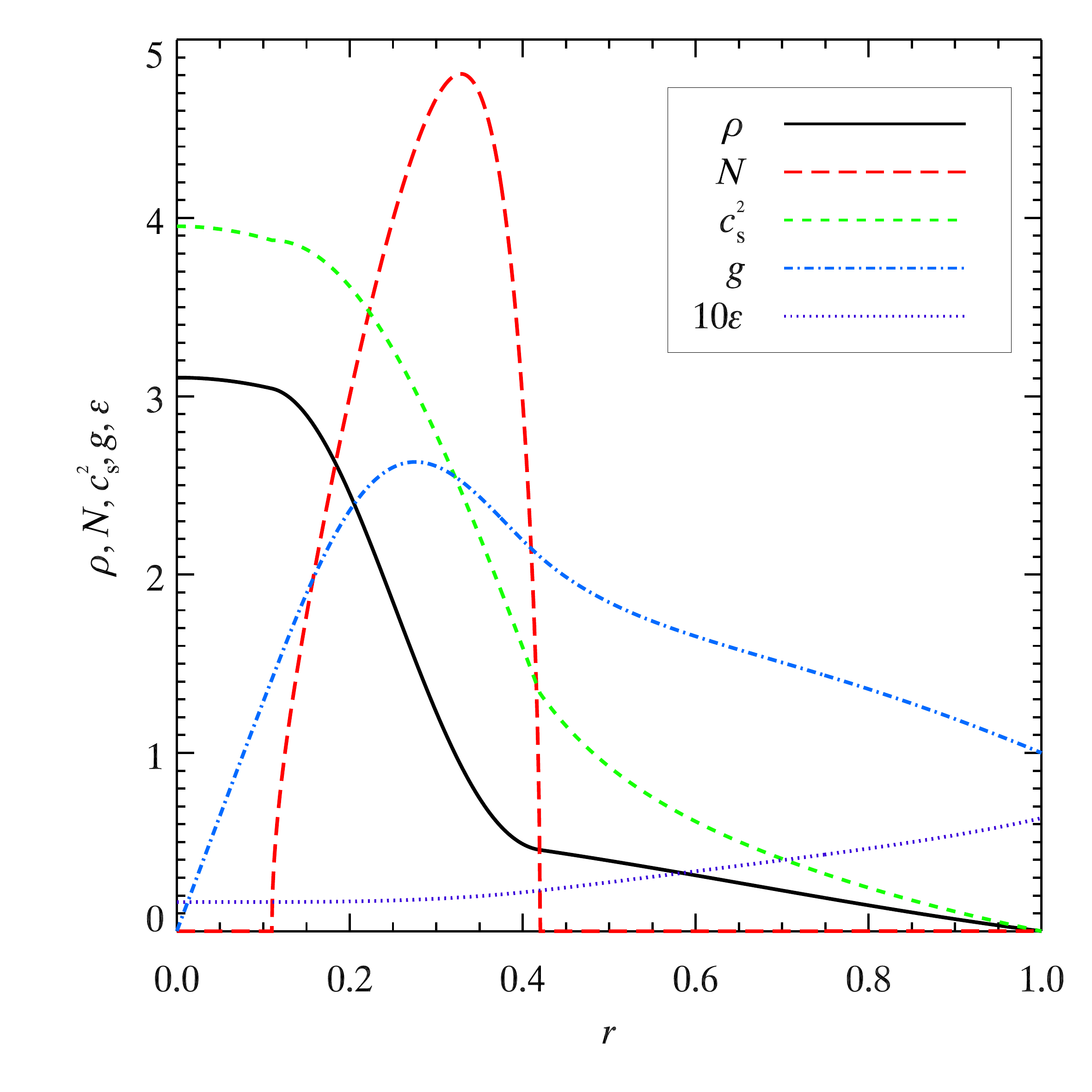}
\end{center} 
\caption{ \label{SatModel} Radial profiles of density $\rho$, sound speed squared $c_s^2$, Brunt-Vaisala frequency $N$, gravitational acceleration $g$, and ellipticity $\varepsilon$ in one of our Saturn models. All quantities are in units with $G=M=R=1$. This model has $R_{\rm eq} = 6.02 \times 10^9$ cm, $R_{\rm po} = 5.46 \times 10^9$ cm, and $J_2=1.6\times10^{-2}$, similar to the measured values of Saturn. For reference, this model has a central density of $9.0 \ {\rm g}/{\rm cm}^3$, and a ``core" mass (defined as the mass within $[r_{\rm in}+r_{\rm out}]/2=0.26$) of $\approx 17 M_\oplus$.}
\end{figure}

After performing the above procedure, we scale the mass and radius of our model to match that of Saturn. From the profiles of $\rho$ and $c_s$ we calculate new values of $g$ and $N$. The resulting $\rho$, $c_s^2$, $N^2$, and $g$ profiles of our reference model are shown in Figure \ref{SatModel}. Note the value of $N^2$ is positive only in the outer core ($r_{\rm in } < r < r_{\rm out}$). Consequently, g-modes are confined to the outer core in our models. This type of model is physically motivated by a scenario in which stable stratification is generated within Saturn by a composition gradient between the rocky/icy inner core and the hydrogen/helium envelope. 

The maximum value of $N$ shown in Figure \ref{SatModel} is very typical for models containing a smooth density profile between the core and the envelope. In the stably stratified region, the Brunt-Vaisala frequency has a value of $N^2 \sim (D/2) g/(r_{\rm out}-r_{\rm in})$. The dense core implied by Saturn's $J_2$ entails typical values of $D\sim 4$ and $g \sim 2.5$, while our models typically have $(r_{\rm out}-r_{\rm in}) \sim r \sim 0.25$. Typical peak values of $N$ in the stably stratified region thus naturally have order of magnitude $N \sim 4$. 

For each model we create, we perturbatively calculate the centrifugal flattening of the planet using Clairaut's equation (see Eggleton 2006). In this approximation, the radius of the spherically symmetric model is perturbed such that
\beq
\label{rac}
r_{\rm ac} = r \big[ 1 - \varepsilon(r) P_2 (\cos \theta) \big],
\eeq
where $r$ is the radius in the spherically symmetric model and $r_{\rm ac}$ is the perturbed radius. The ellipticity of the model, $\varepsilon(r)$, is proportional to the small parameter $(\Omega_s/\Omega_{\rm dyn})^2$, where $\Omega_{\rm dyn} = \sqrt{GM/R^3}$ is the dynamical frequency of Saturn. The details of the calculation are described in \ref{ellipticity}. 

Equation \ref{rac} implies $R_{\rm po} = R[1-\varepsilon(R)]$ and $R_{\rm eq} = R[1+\varepsilon(R)/2]$, which in turn requires
\beq
\label{Rav}
R = R_{\rm po}/3 + 2 R_{\rm eq}/3.
\eeq
The measured polar and equatorial radii of Saturn are $R_{\rm po} = 5.44\times10^9$ cm and $R_{\rm eq} = 6.03\times10^9$ cm. Equation \ref{Rav} then results in $R = 5.83 \times 10^9$ cm. This is the radius of our spherical models. We adopt a spin period of 10 hours, 34 minutes, which entails  $(\Omega_s/\Omega_{\rm dyn})^2 \simeq 0.14$.\footnote{Saturn's spin peroid is not known exactly because the alignment of its magnetic field with its spin axis makes magnetosphere rotation measurements very difficult. However, the adopted spin period is probably accurate to within $\sim 1$\%, which is accurate enough for our purposes since we have excluded higher order rotational effects. See Guillot \& Gautier 2014 for discussion.} Higher order corrections are of order $(\Omega_s/\Omega_{\rm dyn})^4 \simeq 0.019$, thus, we should expect quantities to deviate by a couple percent in a more accurate model.

It is important to note that the value of $\varepsilon(R)$ is somewhat insensitive to the density profile of a reasonable Saturn model. The value of $J_2$, however, can vary considerably. Indeed, the measured value of $J_2$ has led to the conclusion that Saturn must have a core which is much denser than Saturn's average density, and therefore must be enriched in ices, rocks, or metals (Guillot 2005). We construct our models to have $J_2 \simeq 1.6 \times 10^{-2}$ in accordance with the measured value of $J_2$ for Saturn.\footnote{The exact value of $J_2$ will change with the addition of higher order centrifugal corrections, so we do not require our models to have a value of $J_2$ exactly equal to the measured value.} In practice, this means choosing appropriate values of $r_{\rm in}$, $r_{\rm out}$, and $D$ to fulfill this constraint.

\section{Oscillation Modes}
\label{modes}

Saturn rotates rapidly, with $(\Omega_s/\Omega_{\rm dyn}) \simeq 0.38$. The rapid rotation distorts its shape away from sphericity, such that $(R_{\rm eq} - R_{\rm po})/R \simeq 0.1$. These two factors greatly complicate calculations of the oscillation modes of a realistic Saturn model. In this section, we summarize our technique for calculating the mode frequencies and eigenfunctions in the presence of rapid rotation and non-sphericity. The technique utilizes separation of variables in which the angular part of the eigenfunctions is projected onto spherical harmonics, and the radial part is projected onto ``pseudo-modes" which serve as basis functions for computing the normal modes. In the absence of rotation, the pseudo-modes are identical to the orthogonal oscillation modes of a planetary model. Adding rotation modifies the characteristics of the pseudo-modes and introduces new classes of pseudo-modes. Rotation also causes the pseudo-modes to mix with one another via the Coriolis/centrifugal forces. The non-sphericity of Saturn is treated perturbatively, causing additional mixing between the pseudo-modes.

At all times, we consider only linear, adiabatic, fluid oscillations in a uniformly rotating model. The linear and adiabatic approximations are probably very good for the small-amplitude, low wave-number modes we consider. The effect of a solid core was examined by Fuller et al. 2014. Some degree of differential rotation and meridional circulation likely does exist deep within Saturn, however, the amplitude and structure of such flows is not well constrained, and we do not investigate this complication in detail (but see discussion in Section \ref{discussion}). 

The derivation of the pseudo-mode oscillation equations, and our method for solving them, are described in \ref{pseudo}. Including rigid rotation effects to order $(\Omega_s/\Omega_{\rm dyn})^2$ allows pseudo-modes of angular degree $(l,m)$ to mix with pseudo-modes of angular degree $(l\pm2,m)$. We account for this mixing via the methods described in \ref{matrix}. In our formalism, mode eigenfunctions have time and azimuthal dependence
\beq
\bxi \propto e^{i (\omega t + m \phi)} 
\eeq
such that prograde modes have negative values of $m$ for positive mode frequencies $\omega$. 

\begin{figure}
\begin{center}
\includegraphics[scale=0.35]{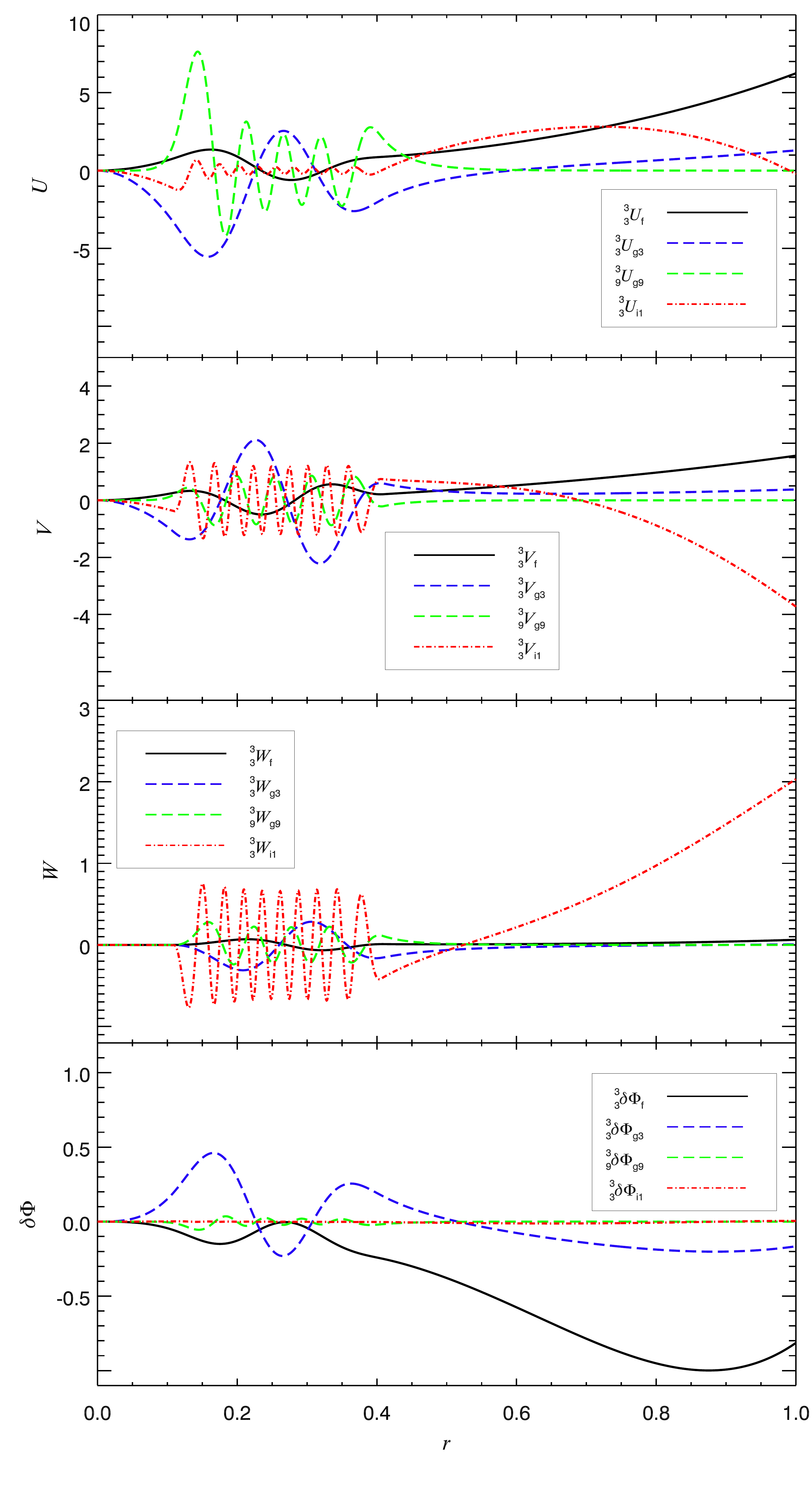}
\end{center} 
\caption{ \label{ModeFunctions} Mode eigenfunctions for an $l=3$, $\omega_\alpha=1.45$ pseudo f-mode (black solid line), $l=3$, $\omega_\alpha=1.36$ pseudo g-mode (blue dashed line), $l=9$, $\omega_\alpha=1.28$ pseudo g-mode (green dashed line), and $l=3$, $\omega_\alpha=0.23$ pseudo i-mode (red dot-dashed line). All are $m=-3$ (prograde) modes, and all quantities are plotted in units of $G=M=R=1$, with pseudo-modes normalized via equation \ref{normpseudo}.  From top to bottom, we plot the radial displacement $U$, horizontal poloidal displacement $V$, horizontal toroidal displacement $W$, and gravitational potential perturbation $\delta \Phi$. The modes are calculated for the planetary model shown in Figure \ref{SatModel}, and correspond to the boxed modes in Figure \ref{Modes}.}
\end{figure}

\subsection{Pseudo-mode Properties}

In the absence of rotation, pseudo-modes are identical to the usual stellar oscillation modes discussed in the literature. There are pseudo p-modes, pseudo f-modes, and in the presence of stratification, pseudo g-modes. Rotation introduces two new classes of pseudo modes: pseudo Rossby modes and pseudo inertial modes. We do not discuss p-modes, as their frequencies are too large to have Lindblad resonances within Saturn's rings. The f-modes, g-modes, and possibly the Rossby modes and inertial modes are important for the problem at hand, and we discuss each of them below.

In the discussion that follows, it is helpful to refer to a dispersion relation. To do this, we use the WKB approximation on equations \ref{eqn1}-\ref{eqn6}, ignoring the gravitational perturbations, and restricting ourselves to the low frequency limit ($\omega^2 \ll L_l^2$, with $L_l^2$ the Lamb frequency). The result is
\beq
\label{dispersion}
k_r^2 = \frac{l^2(l+1)^2}{r^2} \bigg[ \frac{N^2 - \omega^2}{\omega^2} + \frac{q^2 (l+2)^2 S_{lm}^2}{(l+1)(l+2)-mq} \bigg] \bigg[ l(l+1) - mq - \frac{q^2 l^2 (l+2)^2 S_{lm}^2}{(l+1)(l+2)-mq} \bigg]^{-1}.
\eeq
Here, $q=2\Omega_s/\omega$ and $S_{lm}$ is a function of $l$ and $m$ and is of order unity (see \ref{pseudo}). Although this dispersion relation is not very transparent, it is very useful when evaluated in the appropriate limits. In particular, in the non-rotating $q \rightarrow 0$ limit, we obtain the usual g-mode dispersion relation
\beq
\label{gdispersion}
k_r^2 = \frac{l(l+1)(N^2-\omega^2)}{r^2 \omega^2}.
\eeq
Some other useful limits are discussed below.

\begin{figure}
\begin{center}
\includegraphics[scale=0.45]{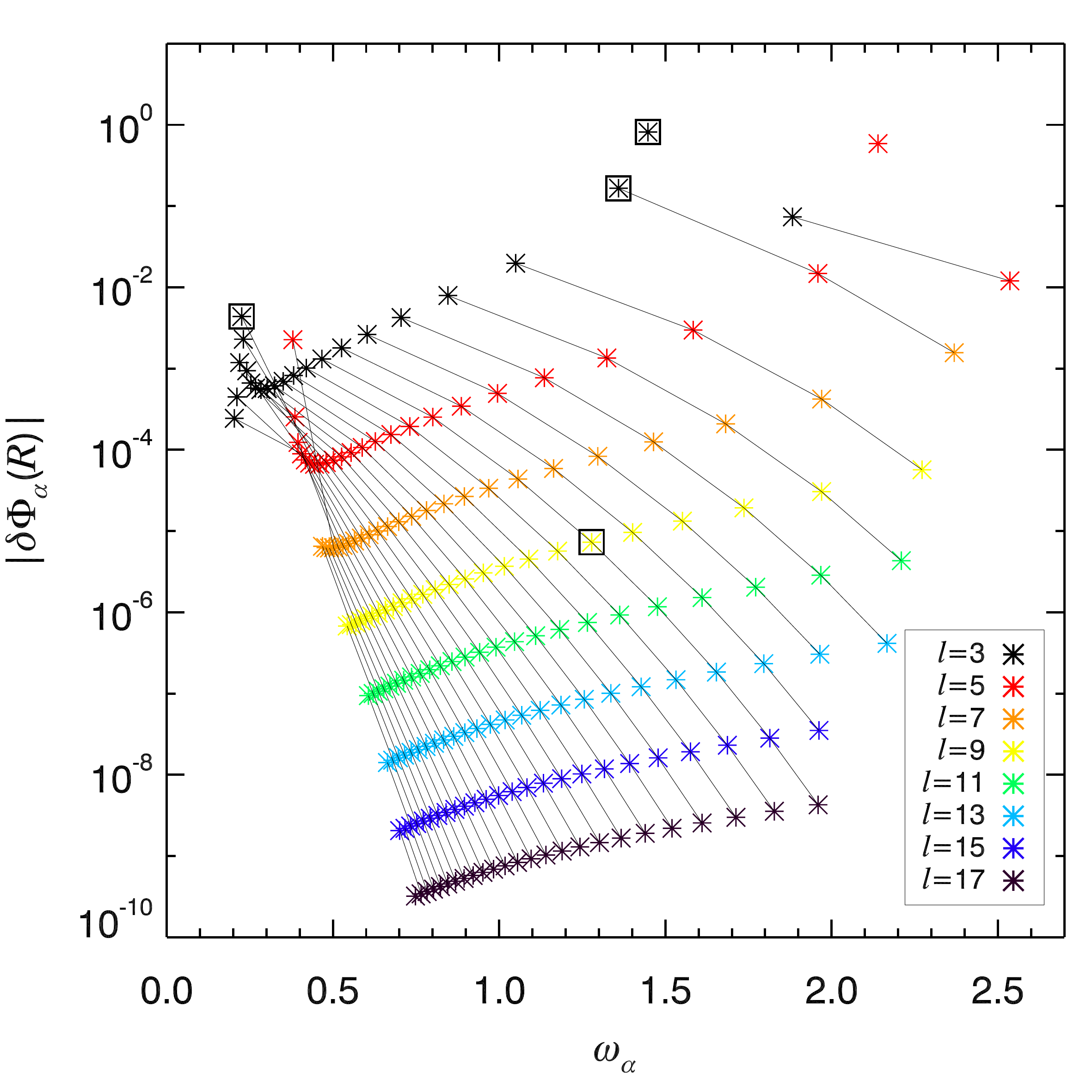}
\end{center} 
\caption{ \label{Modes} Spectrum of prograde $m=-3$ pseudo-modes for the planetary model shown in Figure \ref{SatModel}, with surface gravitational potential perturbations $\delta \Phi_\alpha(R)$ plotted vs. mode frequency $\omega_\alpha$. Quantities are dimensionless and modes are normalized as described in Figure \ref{ModeFunctions}. We have plotted pseudo-modes up to $l=17$, and have only included pseudo-modes with frequencies in the vicinity of the $l=3$ pseudo f-mode. The eigenfunctions of the boxed pseudo-modes are plotted in Figure \ref{ModeFunctions}; the $l=3$ pseudo f-mode is the boxed mode with the largest value of $|\delta \Phi_\alpha(R)|$. Pseudo-modes with the same number of nodes in the g-mode cavity are connected by lines. Mode mixing occurs most strongly between modes of like frequency (i.e., along vertical columns) and between modes of the same radial order $n$ (i.e., along branches connected by lines).}
\end{figure}

\subsubsection{Pseudo f-modes}

In the Saturn models we examine, each angular degree $(l,m)$ family of pseudo-modes typically contains a single f-mode (which are essentially the surface gravity modes discussed in Braviner \& Ogilvie 2014). The f-modes have no nodes in their radial eigenfunctions (in the convective envelope), and have frequencies of approximately
\beq
\label{fmodes}
\omega_{\rm f} \approx \sqrt{l} \Omega_{\rm dyn}.
\eeq
The gravitational potential perturbations of the f-modes are large, and typically $\leftidx{^{\phantom{k} l}_m} \delta \Phi_{\rm f}(R) \gg\leftidx{^{\phantom{k} l}_m} \delta \Phi_{\alpha}(R)$ for all modes $\alpha$ other than the f-mode (under normalization convention of \ref{normpseudo}) . Here, the prescript notation indicates a pseudo-mode of angular dependence $(l,m)$. Because the f-modes have the largest potential perturbations and also have Lindblad resonances within the rings, we expect them to produce the highest amplitude waves in the rings.

Figure \ref{ModeFunctions} displays a plot of the radial displacement $U$, horizontal poloidal component $V$, horizontal toroidal component $W$, and potential perturbation $\delta \Phi$ for the $l=3$ and $m=-3$ pseudo f-mode of the model shown in Figure \ref{SatModel}. The pseudo f-modes are not ``pure" f-modes because they have g-mode character in the g-mode cavity (analogous to the mixed character of oscillation modes in red giant stars), but most of their inertia typically resides in the convective envelope.

\subsubsection{Pseudo g-modes}

Since our models contain a stably stratified region, they support g-modes. The pseudo g-modes are mostly localized to the stably stratified layer, and typically have small gravitational potential perturbations. Figure \ref{ModeFunctions} shows the eigenfunctions of two representative pseudo g-modes.

In the limit $\omega^2 \ll N^2$, pseudo g-modes obey the dispersion relation: 
\beq
\label{gdisp}
k_r^2 = \frac{l^2(l+1)^2}{r^2} \bigg[ \frac{N^2}{\omega^2} + \frac{q^2 (l+2)^2 S_{lm}^2}{(l+1)(l+2)-mq} \bigg] \bigg[ l(l+1) - mq - \frac{q^2 l^2 (l+2)^2 S_{lm}^2}{(l+1)(l+2)-mq} \bigg]^{-1}.
\eeq
Note that g-modes (of the same radial order) have higher frequencies for larger values of $N$, which can be produced by larger density gradients in a given model. G-modes also have higher frequencies for g-mode cavities at small radii $r$, and for large values $l$.  

The terms containing $q$ in equation \ref{gdisp} modify the pseudo g-modes from the non-rotating g-mode dispersion relation of equation \ref{gdispersion}. For $q\gtrsim1$ they typically increase the radial wavenumber of pseudo-gmodes. When the second term in brackets is equal to zero, the wavenumber diverges. This divergence occurs occurs at $\omega < \Omega_s$ for $l = |m|$, and at $\omega \simeq \Omega_s$ for $l \gg |m|$. For even lower frequencies, the second term in brackets is negative, such that there are no very low frequency pseudo g-modes. However, this does not imply that low frequency g-modes do not exist in rapidly rotating stars/planets, as the pseudo-mode mixing process (see Section \ref{mix}) allows the normal modes to have frequencies different from the pseudo-modes. We therefore caution that our method for calculating (and understanding) very low frequency modes may not be optimal. 

In the super-inertial regime of interest ($\omega>2\Omega_s$), equation \ref{gdispersion} translates to the approximate g-mode spectrum:
\beq
\label{gdispersion2}
\leftidx{^{l}} \omega_n \sim \frac{\sqrt{l(l+1)} N }{\pi n}
\eeq
where $n$ is the number of nodes in the radial eigenfunction. We find that in our Saturn models, in which the value of $N$ is peaked in the outer core at at values $N \sim 5$, the spectrum of pseudo g-modes typically extends to {\it higher} frequencies than the f-modes. Low $l$ pseudo g-modes near the low $l$ pseudo f-modes typically have $n\sim 3$. Higher $l$ pseudo g-modes have even larger frequencies, meaning that their spectrum is dense in the vicinity of the low $l$ pseudo f-modes. Figure \ref{Modes} shows the pseudo mode spectrum up to $l=17$. We can see that our models contain many pseudo g-modes in the same frequency regime as low $l$ pseudo f-modes. 

Equation \ref{gdispersion2} can also be used to find g-mode frequency spacings:
\beq
\label{omspacen}
\leftidx{^{l}} \Delta \omega_n \sim \leftidx{^{l}} \omega_n \frac{\Delta n}{n}
\eeq
and 
\beq
\label{omspacel}
\leftidx{^{l}} \Delta \omega_n \sim \leftidx{^{l}} \omega_n \frac{\Delta l}{l}.
\eeq
Low $l$ pseudo g-modes of consecutive radial order have frequency spacings of $\leftidx{^{l}}\Delta \omega_n \sim \leftidx{^{l}} \omega_n/3$ at frequencies near the low $l$ pseudo f-modes. pseudo g-modes of the same radial order $n$ but with angular number varying by $\Delta l=2$ have $\leftidx{^{l}} \Delta \omega_n \sim 2 \ \leftidx{^{l}} \omega_n/l$. These frequency spacings are important for calculating the probability of finding g-modes with frequencies near the f-modes (see Section \ref{results}) and for understanding how modes can mix up to high values of $l$ (see Section \ref{mix}). 


\subsubsection{Pseudo Inertial Modes}

In addition to f-modes and g-modes, our rotating models support inertial modes. Whereas f-modes and g-modes are restored by self-gravity and buoyancy, respectively, inertial modes are restored by the Coriolis force. The pseudo inertial modes are restricted to the sub-inertial frequency regime $\omega < 2 \Omega_s$. Like g-modes, they exhibit small pressure and gravitational perturbations. Figure \ref{ModeFunctions} shows the eigenfunction of an inertial pseudo-mode. Like the f-modes, the inertial modes are mostly restricted to the convective envelope, although they may have g-mode character within the g-mode cavity.

The inertial pseudo mode dispersion relation can be found by taking the $N=0$ limit of equation \ref{dispersion}:
\beq
\label{idispersion}
k_r^2 = \frac{l^2(l+1)^2}{r^2} \frac{-(l+1)(l+2) + m q + q^2 (l+2)^2 S_{lm}^2}{\big[m^2-l^2(l+2)^2 S_{lm}^2 \big]q^2 - 2m(l+1)^2 q + l(l+1)^2(l+2)}.
\eeq
This dispersion relation is notable because it depends only on the frequency ratio $q$, and the values of $l$ and $m$, without any dependence on the material properties of the planet.\footnote{Inertial modes can be shown to have a dispersion relation $\omega/(2 \Omega_s) = {\hat {\bf z}} \cdot {\bf k} / k$ (see, e.g., Greenspan 1968). Like equation \ref{idispersion}, it shows that inertial modes are sensitive only to the spin frequency and their angle of propagation.} Note that the radial wavenumber diverges when the denominator is equal to zero, as discussed above. It is possible that the divergence of the wave vector of the pseudo modes is related to the wave number divergence of inertial modes at critical latitudes (discussed in Ogilvie \& Lin 2004 and Wu 2005) although we do not explicitly examine this phenomenon here.

Because the pseudo inertial modes are restricted to the sub inertial frequency range, they all have lower frequency than Saturn's f-modes, and cannot undergo degenerate mixing (avoided crossings) with them. For this reason, we do not perform a detailed investigation of pseudo inertial modes. We note, however, that the ``fundamental" (only one node in its radial eigenfunction) pseudo inertial mode shown in Figure \ref{ModeFunctions} mixes relatively strongly with the f-modes because they are both localized primarily in the convective envelope. Finally, we comment that our technique for solving for normal modes from a pseudo mode basis (see \ref{matrix}) may not be tractable for solving for inertial modes, as an accurate calculation would likely require the inclusion of pseudo inertial modes of very high radial wave number $n$ and angular degree $l$. It may be simpler to use the techniques discussed in Ogilvie \& Lin 2004 and Wu 2005.

\subsubsection{Pseudo Rossby Modes}

Rotation also introduces a class of modes not often explored in the literature, which we refer to as pseudo Rossby modes (these modes have been explored using the traditional approximation in Lee \& Saio 1997, Fuller \& Lai 2014). These modes require stable stratification to exist, and are thus similar to g-modes. However, the Rossby modes exist at lower frequency, and are purely {\it retrograde} modes. The Rossby pseudo-modes exist in the $\omega\rightarrow0$, $q \rightarrow \infty$ limit of equation \ref{dispersion}. The result is 
\beq
\label{rdispersion}
k_r^2 = \frac{N^2}{2r^2 \omega \Omega_s} \frac{l^2(l+1)^2 m}{l^2(l+2)^2 S_{lm}^2 - m^2}.
\eeq
The denominator of the fraction on the right is always positive, thus, for positive frequencies, we require $m>0$ to have $k_r^2 >0$ (i.e., only retrograde modes can exist). 

The pseudo Rossby modes are retrograde modes, and thus cannot undergo avoided crossings with the prograde modes we are interested in. However, because they have similar characteristics to g-modes, they mix somewhat strongly with them. Moreover, because their frequencies are small ($\omega \ll \Omega_s$), the rotational mixing between pseudo Rossby modes of different $l$ is quite strong. Like the inertial modes, we do not include a more detailed analysis of the Rossby modes since they do not exist in the super inertial regime.

\subsubsection{Mixed Modes}

In reality, the pseudo-modes are not strictly f-modes or g-modes or inertial modes, but share qualities of each. In the Saturn model shown in Figure \ref{SatModel}, the $l=3$ pseudo g-mode from Figure \ref{ModeFunctions} exhibits some f-mode character due to its proximity in frequency to the $l=3$ pseudo f-mode, giving it an enhanced surface potential perturbation $\delta \Phi(R)$. In turn, the $l=3$ pseudo f-mode has some g-mode character, causing its eigenfunction to have appear wave-like in the g-mode cavity. Additionally, the inertial mode from Figure \ref{ModeFunctions} has some g-mode character. The latter mixing occurs because prograde pseudo modes in the sub-inertial frequency range behave like g-modes in stably stratified regions, and inertial modes in convective regions.\footnote{For retrograde modes, this is not the case because the gravity and inertial pseudo-mode frequency regimes are distinct.} This feature is responsible for the spikes in $\delta \Phi(R)$ in Figure \ref{Modes} at $\omega_\alpha \approx 0.3$, which are created by gravito-inertial mixed modes near the ``fundamental" pseudo inertial mode (the inertial mode with only one node in its eigenfunction in the convective envelope).

\section{Mode Mixing}
\label{mix}

As discussed above, the pseudo-modes are not normal oscillation modes because they mix with one another due to the Coriolis and centrifugal forces, as well as the elliptical shape of the planet. A useful analogy to understand planetary rotational mode mixing is the mode mixing between atomic electron energy levels induced by an electric field (the Stark effect) or a magnetic field (the Zeeman effect). In the limit of a weak magnetic field, the Zeeman effect splits degenerate energy levels into a multiplet. The observed frequency splitting can then be used to infer the strength of the magnetic field. Similarly, in slowly rotating stars or planets, the Coriolis force induces a small frequency splitting which can be used to measure the rotation rate.

However, for a strong magnetic field, atomic energy levels are split by so much that they overlap with adjacent energy levels. The energy levels of different electron wave functions cannot cross (they cannot be exactly degenerate); instead they undergo an avoided crossing in which the modes mix with one another, exchanging character. Near the avoided crossing the resulting electron wave functions are mixtures (superpositions) of the non-perturbed wave functions, and the energy levels (eigenfrequencies) of the states are similar but not exactly the same. A very similar process occurs in rapidly rotating stars and planets. In our case, the rapid rotation of Saturn perturbs f-mode eigenfrequencies enough that they overlap with neighboring g-mode eigenfrequencies. The f-modes and g-modes mix with one another, forming mixed modes that are superpositions of the f-mode and g-modes, and whose frequencies are similar but not equal.

The dynamics of rotational mode mixing is discussed extensively in Fuller et al. 2014, here we review the basic physics. The mode mixing angle between two pseudo-modes $\alpha$ and $\beta$ is
\beq
\tan (2\theta_{\alpha \beta}) = \frac{C_{\alpha \beta}}{\omega_\alpha^{(1)} - \omega_\beta^{(1)}}.
\eeq
Here $C_{\alpha \beta}$ is the mode mixing coefficient, while $\omega_\alpha^{(1)}$ and $\omega_\beta^{(1)}$ are the perturbed mode frequencies $\omega_\alpha^{(1)} = \omega_\alpha + C_{\alpha \alpha}$. The effective value of $C_{\alpha \beta}$ is a combination of the mixing coefficients $W_{\alpha,\beta}$ (due to the Coriolis force), $\delta V_{\alpha \beta}$ (due to centrifugal/ellipticity perturbations to the potential energy) and $\delta T_{\alpha \beta}$ (due to ellipticity perturbations to the kinetic energy). These mixing coefficients  are discussed in more detail in \ref{matrix}, and expressed explicitly in DT98. 

Strong mode mixing occurs when $|\omega_\alpha^{(1)} - \omega_\beta^{(1)}| < C_{\alpha \beta}$. The simplest way for this to occur is for two modes to be nearly degenerate, i.e., for $\omega_\alpha^{(1)} \simeq \omega_\beta^{(1)}$. We refer to this case as degenerate mode mixing, which is equivalent to saying two modes are undergoing an avoided crossing. However, strong mode mixing can occur at larger frequency separations if the value of $|C_{\alpha \beta}|$ is large. This typically only occurs for modes of the same type (g-mode, inertial mode, etc.) and with the same number of nodes in their radial eigenfunctions. 

Mixed modes have eigenfunctions that are superpositions of the original pseudo-modes. A pseudo g-mode mixed with a pseudo f-mode obtains a contribution of $\sim \sin(\theta_{\alpha {\rm f}}) \bxi_{\rm f}$ to its eigenfunction (where $\bxi_{\rm f}$ is the pseudo f-mode eigenfunction). Since g-modes typically have very small intrinsic potential perturbations, their potential pertrubations are often dominated by the pseudo f-mode contribution such that they have $\delta \Phi_\alpha(R) \approx \sin(\theta_{\alpha {\rm f}}) \delta \Phi_{\rm f}(R)$. 

Our techniques only allow for mixing between pseudo-modes of the same $(l,m)$, and pseudo-modes of $(l\pm2,m)$. However, pseudo-modes with larger differences in $l$ may still mix indirectly through intermediary modes, as was discussed in Fuller et al. 2014. As the number of intermediary pseudo-modes increases, we may expect that strong mixing becomes more difficult, although it remains possible. So, while pseudo g-modes of large $l$ can mix with the low $l$ pseudo f-modes, they need to be very close in frequency for strong mixing to occur. This idea motivates our proposal that the small frequency differences between the observed waves in Saturn's rings are due to mode mixing between the low $l$ pseudo f-modes and larger $l$ pseudo g-modes. 

We do not attempt to provide a detailed investigation of all the different types of mixing (Coriolis, potential, kinetic) between all the different types of pseudo-modes (f-modes, g-modes, inertial modes, Rossby modes). Here we only comment on the order of magnitude of the mixing. The maximum possible values of the mixing coefficients are of order $W_{\rm max} \sim \Omega_s$, $\delta V_{\rm max} \sim \Omega_s^2$, and $\delta T_{\rm max} \sim \varepsilon$. Typical values are even smaller, especially for modes whose radial orders differ, or which propagate in different cavities. Low $l$ f-modes have largest mixing coefficients with neighboring g-modes of the same $l$, and with the fundamental inertial mode. The g-modes usually mix most strongly with g-modes of $l \pm 2$ with the same radial order $n$. 

We can now understand how mixing between the pseudo-modes shown in Figure \ref{Modes} will proceed. The pseudo g-modes will tend to mix with g-modes of the same radial order, i.e., along the g-mode branches connected by lines in Figure \ref{Modes}. The pseudo modes will also mix with modes of nearly the same frequency, i.e., along vertical columns in Figure \ref{Modes}. We also note that pseudo g-modes can mix with their negative frequency (retrograde) and Rossby mode counterparts. However, since these modes are further away in frequency, the degree of mixing should be smaller.

Although the pseudo-modes do not experience any self-mixing due to the Coriolis force\footnote{The Coriolis self-mixing coefficient $W_{\alpha \alpha}$ of modes calculated for a non-rotating model is identical to the frequently discussed rotational frequency perturbation $\delta \omega$ that is valid in the slow rotation $(\Omega_s \ll \omega)$ limit. As a check on our numerical code, we confirm that the frequencies of the pseudo-modes are shifted by $\delta \omega$ compared to modes calculated for a non-rotating model, in the limit of slow rotation.}, they do experience self-mixing due to the potential $\delta V_{\alpha \alpha}$ and kinetic $\delta T_{\alpha \alpha}$ mixing coefficients, causing their frequencies to shift. The largest frequency shifts occur for the pseudo f-modes, and are due to the centrifugal component of $\delta V_{\alpha \alpha}$. This causes the computed normal f-modes to have frequencies roughly $0.1 \Omega_{\rm dyn}$ lower than their pseudo-mode counterparts. The centrifugal correction is thus crucial for obtaining f-mode frequencies within $\sim 1 \%$ of the observed values.


\section{Results}
\label{results}

Our ultimate goal is to compare our computed oscillation mode spectrum to the observed mode spectrum. Unfortunately, the observed spectrum is incomplete, as only modes with Lindblad resonances that lie within the rings can excite observable waves. Moreover, only modes with large enough gravitational potential perturbations to subtantially affect the rings can be observed. This likely means that only f-modes and g-modes undergoing avoided crossings with the f-modes can be observed. We compare our results with the six waves observed by HN13, keeping in mind that the observed mode spectrum is incomplete. 

To estimate mode amplitudes, we use the same technique adopted in Fuller et al. 2014 (see Section 5 and Appendix D of that work). The important idea is that the optical depth variation $\delta \tau$ of the ring waves is proportional to the amplitude $A_\alpha$ and gravitational potential perturbation $\delta \Phi_\alpha(R)$ of mode driving them. If one can estimate the value of $\delta \Phi_\alpha(R)$ of the mode, one can use the measured optical depth variation to calculate the mode amplitude. To do this, we assume the most prominent $m=-3$ wave is generated by a mode with the same value of $\delta \Phi_\alpha(R)$ as the $l=3$, $m=-3$ pseudo f-mode of our model. We then calculate the mode frequency$\leftidx{^{\phantom{-} 3}_{-3}}\omega$ and amplitude $|\leftidx{^{\phantom{-} 3}_{-3}} A|$ required to generate this wave, the latter of which is calculated from equation D.7 of Fuller et al. 2014. Each mode amplitude of our model is then calculated assuming energy equipartition, i.e., $|A_{\alpha}|^2 \omega_\alpha^2 = |\leftidx{^{\phantom{-} 3}_{-3}} A|^2 \leftidx{^{\phantom{-} 3}_{-3}}\omega^2$. We re-examine the validity of energy equipartion in Section \ref{discussion}. Finally, we caution that the amplitude $|\leftidx{^{\phantom{-} 3}_{-3}} A|$ inferred from the ring observations is only approximate, as the precise amplitude of the waves in the rings is somewhat difficult to measure because of non-linear and wave damping effects.

In what follows, it is helpful to recall the relation between rotating frame mode frequency $\omega_\alpha$, inertial frame mode frequency $\sigma_\alpha$, and the wave pattern frequency $\Omega_{p,\alpha}$:
\beq
\omega_\alpha - m \Omega_s = \sigma_\alpha = -m \Omega_{p,\alpha}.
\eeq

\begin{figure}
\begin{center}
\includegraphics[scale=0.4]{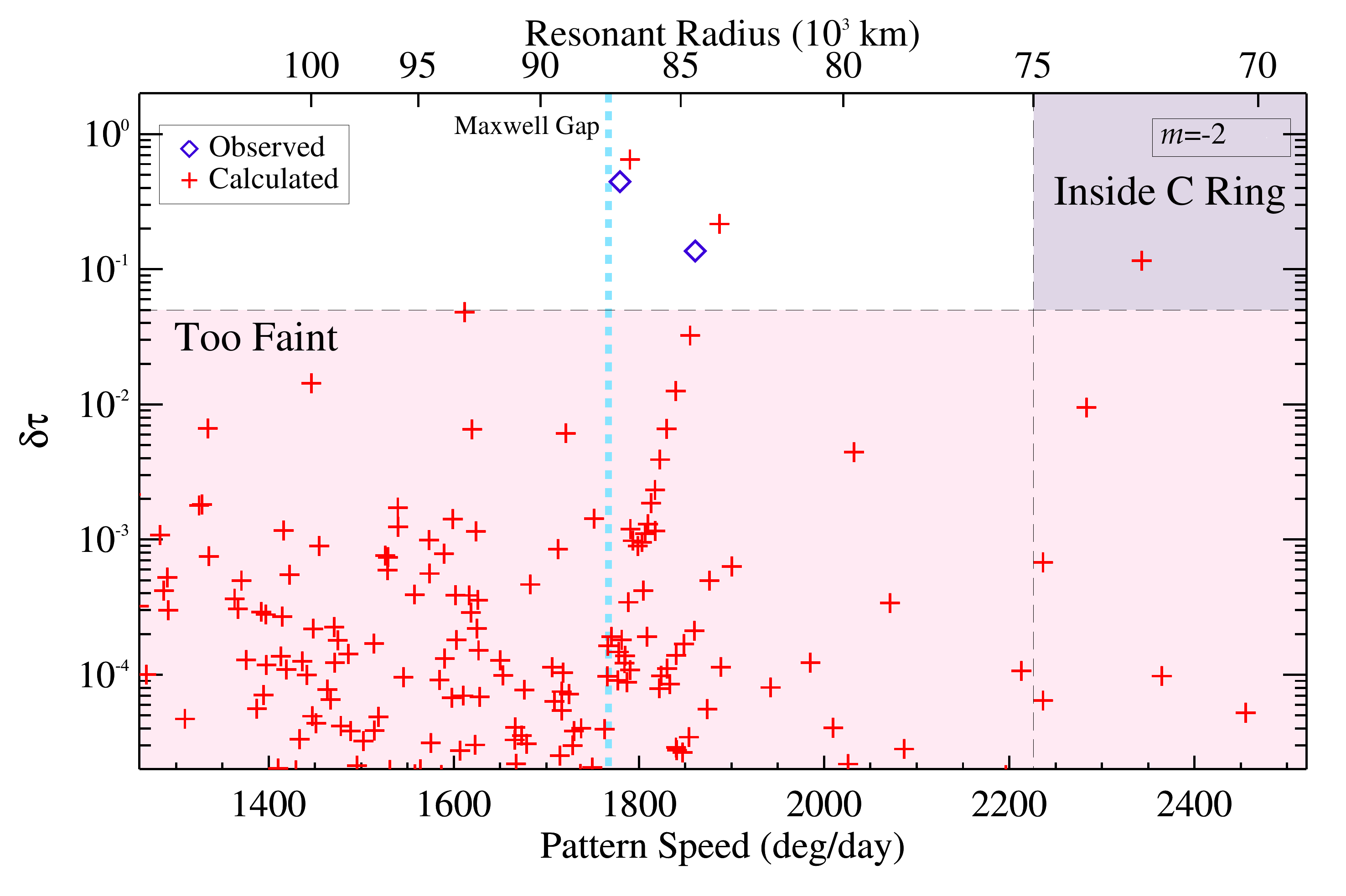}
\end{center} 
\caption{ \label{m2} Wave amplitudes in Saturn's rings (characterized by their maximum optical depth variation $\tau$) as a function of wave pattern speed $\Omega_p = -\sigma/m$, for $m=-2$ waves. The blue diamonds are the waves measured by HN13, while the red crosses are waves generated by the normal oscillation modes we have computed. The modes are calculated for the Saturn model shown in Figure \ref{SatModel}. Modes in the pink shaded region are not observed because their gravitational perturbations are too weak to generate detectable waves in the rings. Modes in the purple shaded region have Lindblad resonances inside the inner edge of the C-ring where it is difficult to detect waves. The vertical blue dashed line denotes the location of the Maxwell gap, which may be opened by the $l=2$, $m=-2$ f-mode (see Section \ref{discussion}).}
\end{figure}

\begin{figure}
\begin{center}
\includegraphics[scale=0.4]{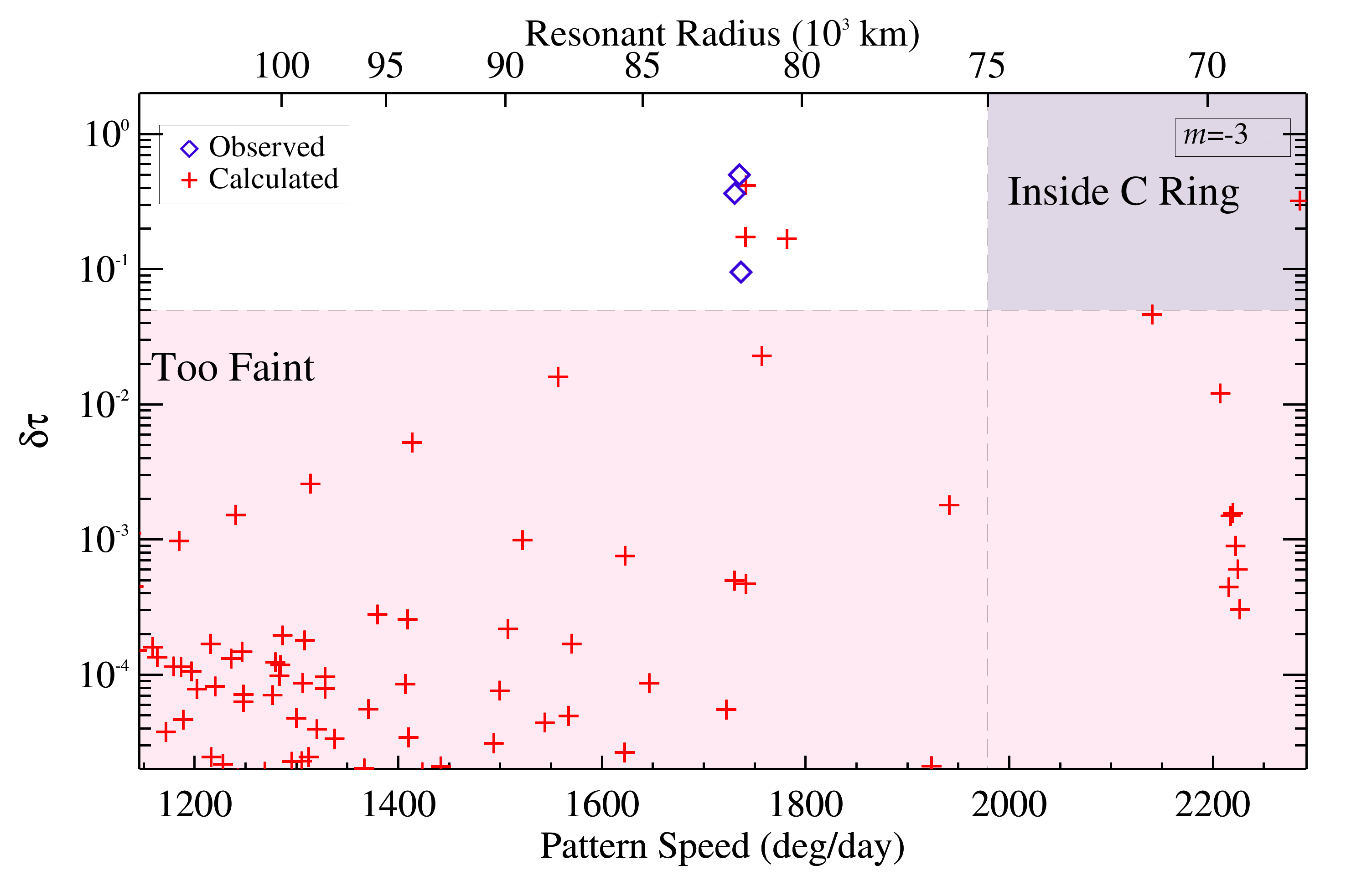}
\end{center} 
\caption{ \label{m3} Same as Figure \ref{m2}, but for $m=-3$ modes. }
\end{figure}

\begin{figure}
\begin{center}
\includegraphics[scale=0.4]{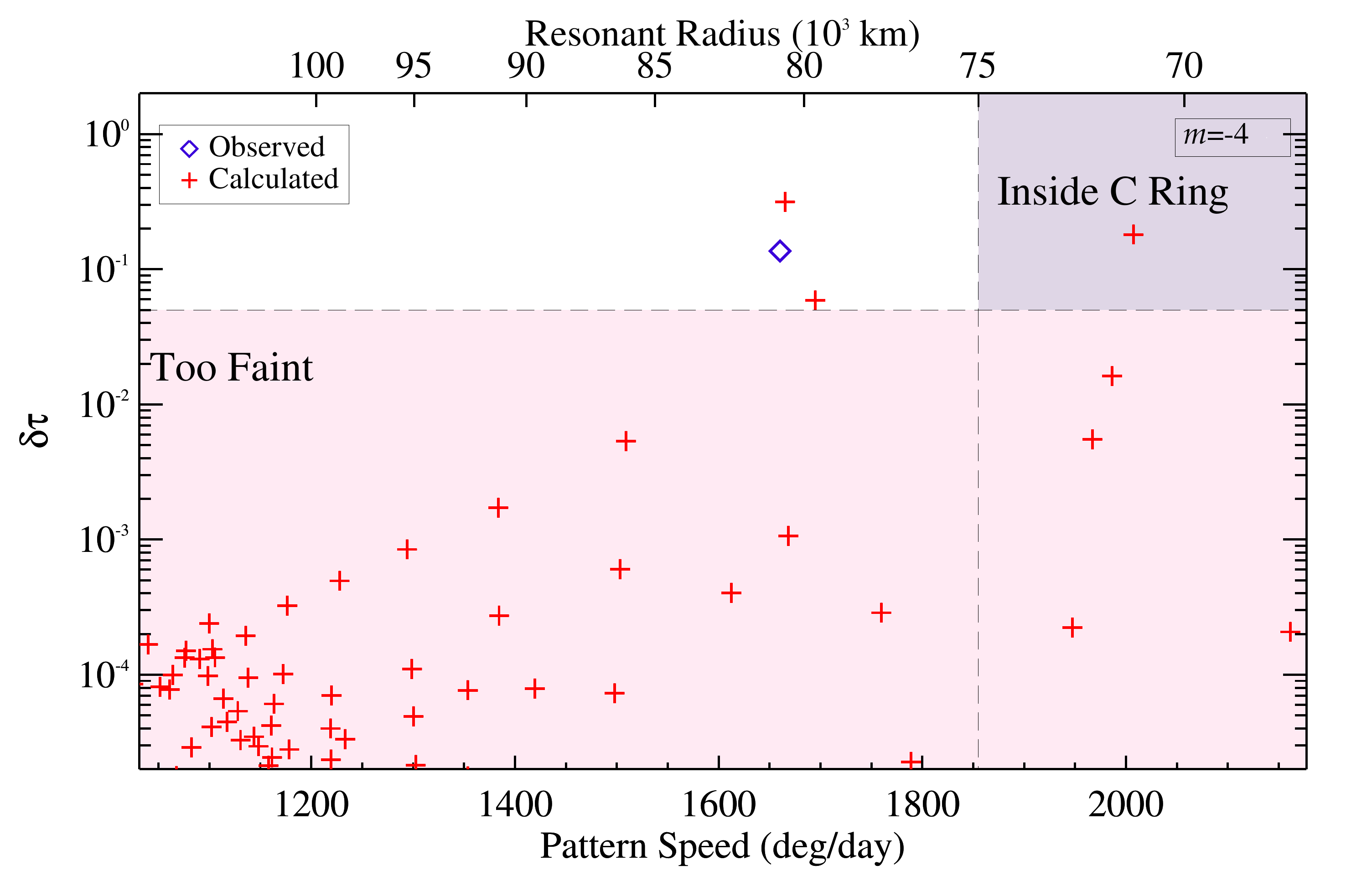}
\end{center} 
\caption{ \label{m4} Same as Figure \ref{m2}, but for $m=-4$ modes. }
\end{figure}

\begin{figure}
\begin{center}
\includegraphics[scale=0.4]{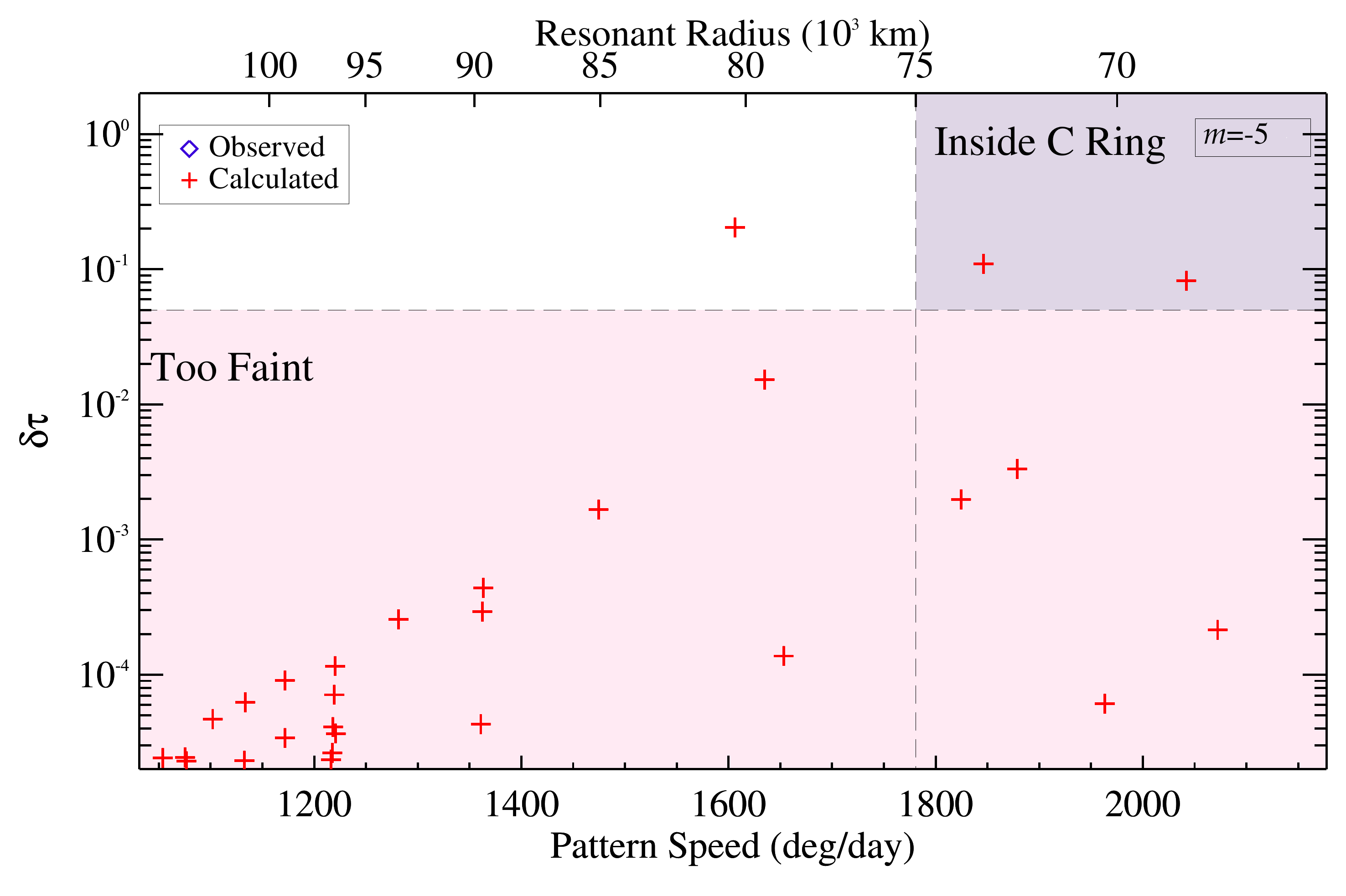}
\end{center} 
\caption{ \label{m5} Same as Figure \ref{m2}, but for $m=-5$ modes. No $m=-5$ waves have been detected, likely because the mode amplitudes are too small (see Section \ref{discussion}).}
\end{figure}

We begin by examining Figure \ref{m2} and Table \ref{table}. The two observed waves (Waves A and B in Table \ref{table}) have pattern frequencies near the expected location for $l=2$, $m=-2$ f-modes (M91, MP93), and are separated by $\sim 5 \%$ in frequency. Our computed normal mode spectrum also contains two modes within $\sim 2 \%$ of these observed pattern speeds. For comparison, the frequencies $\sigma_\alpha$ of these two modes are listed in Table \ref{table} in rows A and B. The wave amplitude ratio of our two modes also approximately matches the observed amplitude ratio. In our analysis, these two modes are the $l=2$, $m=-2$ f-mode, and a neighboring $l=2$, $m=-2$, $n=3$ g-mode (i.e., the $\leftidx{^{\phantom{-} 2}_{-2}}\omega_{g3}$ mode). The g-mode obtains a large potential perturbation due to mixing with the f-mode, allowing its effect on the rings to be observed. Both modes are mixed with other values of $l$ but are dominated by their $l=2$ components. 

\begin{figure}
\begin{center}
\includegraphics[scale=0.45]{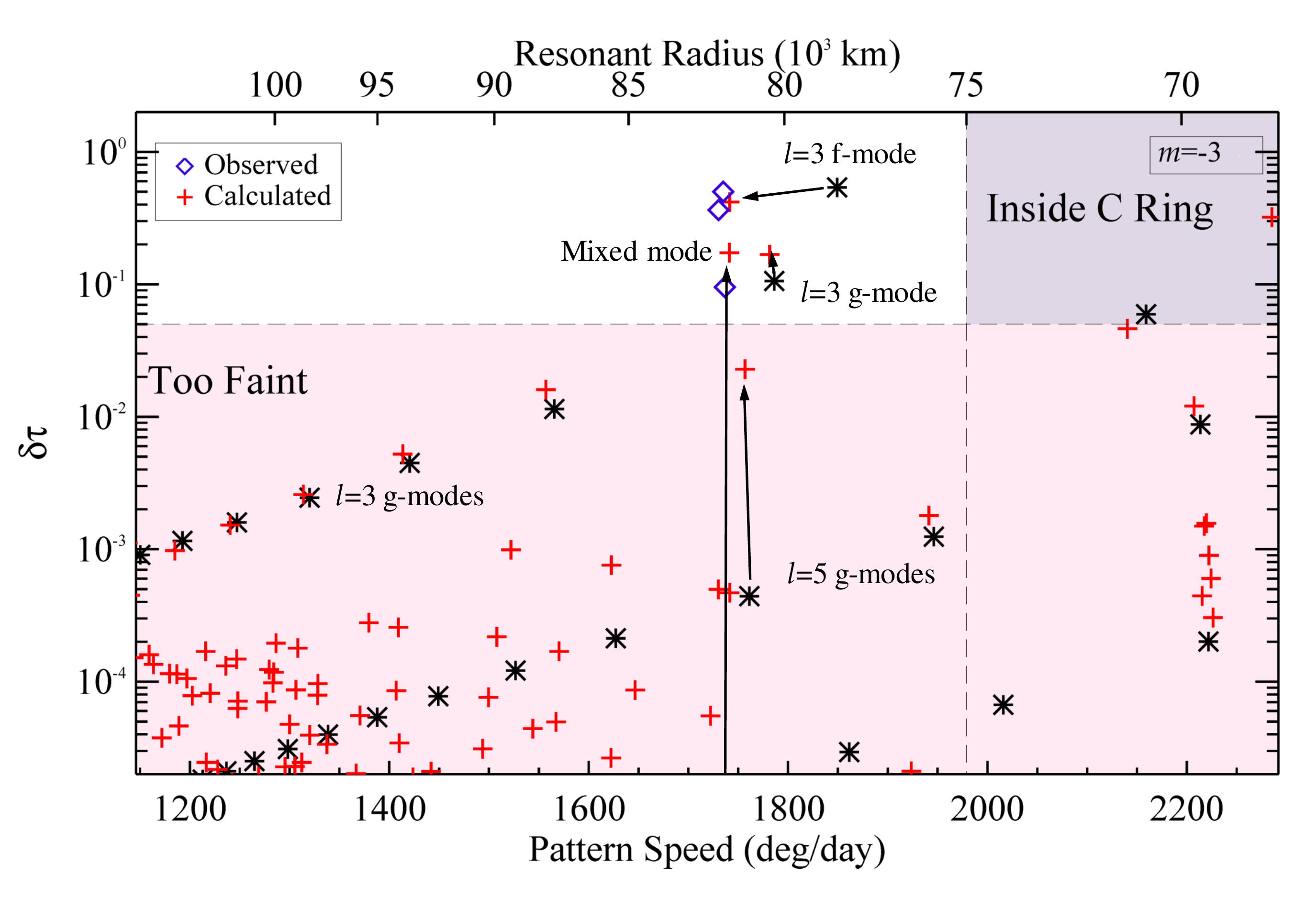}
\end{center} 
\caption{ \label{m3fancy} Same as Figure \ref{m3}. In this version, we have included the locations of the pseudo-modes (black asterisks). Where appropriate, we have drawn arrows from the location of the pseudo-modes to the normal modes, indicating which pseudo-modes are the dominant component of each normal mode. The arrow extending from beneath the plot originates from an $l=7$ pseudo g-mode just below the bottom of the plot.}
\end{figure}

Our Saturn model was slightly fine-tuned to generate a frequency splitting of $\sim 5 \%$ between the largest amplitude $m=-2$ modes. However, spitting of this magnitude is very common. Figures \ref{m3} and \ref{m4} show very similar splitting, in which there is a $\leftidx{^{\phantom{-} l}_{-l}}\omega$ g-mode with frequency split by a few percent from the f-mode frequency. The models need not be fine-tuned to qualitatively reproduce this feature. The reason for such splitting is as follows. 

\begin{table*}
\begin{center}
\caption{\label{table} Observed wave properies from HN13. $|\delta \tau|$ is the approximate maximum semi-amplitude
  of the optical depth variation associated with each wave. $\sigma_{\rm ob}$ is the angular mode frequency (in the inertial frame) of the mode driving each wave. The values of $\sigma_{\rm ca}$ are our calculated frequencies of modes which would produce visible density waves. These modes correspond to the red crosses in Figures \ref{m2}-\ref{m4} which are closest to the observed waves, and which are discussed in the text.}
\begin{tabular}{@{}ccccccc}
\hline\hline
Wave & $m$ & Resonant location & $|\delta \tau|$ & $\Omega_p$ & $\sigma_{\rm ob}$ & $\sigma_{\rm ca}$ \\
 &  & (km) & & (deg/day) & ($\mu$Hz) & ($\mu$Hz)\\
\hline
A & -2 & 87189 & $0.14$ & 1779.5 & 718.94 & 728.37  \\
\hline
B & -2 & 84644 & $0.09$ & 1860.8 & 751.78 & 767.93 \\
\hline
C & -3 & 82209 & $0.15$ & 1730.3 & 1048.6 & 1062.4 \\
\hline
D & -3 & 82061 & $0.21$ & 1735.0 & 1051.4 & 1062.5 \\
\hline
E & -3 & 82010 & $0.07$ & 1736.6 & 1052.4 & 1087.2 \\
\hline
F & -4 & 80988 & $0.09$ & 1660.3 & 1341.6 & 1354.4 \\
\hline\hline
\end{tabular}
\end{center}
\end{table*}

Equation \ref{omspacen} demonstrates that the frequency spacing between $l=|m|$ g-modes in the vicinity of the low $l=|m|$ f-modes is of order $\omega_{\rm f}/3$. The average spacing between an f-mode and the nearest g-mode is one quarter of the g-mode frequency spacing, i.e., $\sim \omega_{\rm f}/12$. The fractional frequency spacing of $\sigma_{\rm f}$ is smaller because the prograde modes are all shifted to higher frequencies in the rotating frame, such that typical frequency spacings are of order $\sim \sigma_{\rm f}/20$. It is thus quite common to have a g-mode whose frequency is within several percent of the f-mode.

In our models, the g-modes are more strongly mixed with the f-modes for smaller values of $l$. The reason is that higher $l$ f-modes are confined closer to the surface of the planet, away from the g-mode cavity deep in the interior. Therefore, low $l$ f-modes mix more strongly with the g-modes, which may explain why a frequency splitting of $\sim 5 \%$ is only observed near the $l=2$, $m=-2$ f-mode. Such splitting also occurs in Figures \ref{m3} and \ref{m4}, but only because this particular Saturn model happens to have $l=3$ and $l=4$ g-modes whose frequencies are closer than average to the f-mode frequencies. Nonetheless, it is very possible that there exist additional waves near the observed $m=-3$ and $m=-4$ waves, which have smaller amplitudes, and which are split in frequency by a few percent. 

Let us now turn our attention to Figure \ref{m3fancy}. The three observed $m=-3$ waves have frequencies that are split an order of magnitude less than the $m=-2$ waves, i.e., their frequencies are split by less than a percent. Such splitting cannot be explained by mode mixing between f-modes and neighboring g-modes of the same $l$. Instead, the splitting may be due to avoided crossings between the f-mode and g-modes of much higher values of $l$. These avoided crossings could be very common, because the g-mode spectrum near the f-mode becomes very dense as we proceed to large values of $l$ (see Figure \ref{Modes}). Figure \ref{m3} shows that there is a mixed mode very near the f-mode, split by $\sim 0.01 \%$ in frequency. The frequencies $\sigma_\alpha$ of these two modes are listed in Table \ref{table} in rows C and D (the mode in row E is the $l=3$ g-mode discussed above). In this case, the fine-splitting arises due to an avoided crossing between the $l=3$ f-mode and an $l=7$ g-mode.

However, we find that strong mixing between the $l=|m|$ f-modes and high $l$ g-modes is uncommon. It appears that the effective mixing coefficients are small, requiring very small frequency separations $\omega_\alpha^{(1)} - \omega_\beta^{(1)}$ for strong mixing to occur. The strong mixing shown in Figures \ref{m3fancy} occurs because the model has been fine-tuned to produce a resonance between the $l=3$ f-mode and an $l=7$ g-mode. We find such strong mixing only occurs in about one tenth of our planet models. Simultaneous strong mixing between three modes, as appears to be observed, is even less common. It is thus difficult to reconcile the observations with the notion of strong mixing between low $l$ f-modes and high $l$ g-modes, indicating that some un-included physical effects may be important (see Section \ref{discussion}). 

Finally, let us examine Figures \ref{m4} and \ref{m5}. The $l=4$ and $l=5$ f-modes have higher frequencies, where the g-mode spectrum is less dense and where mode mixing coefficients are smaller. It is therefore not surprising that no fine splitting is observed for the $m=-4$ mode. The $l=5$, $m=-5$ f-mode is not observed at all, even though our calculations suggest it should be observable. This indicates that modes of higher $l$ and/or higher frequency are excited to lower amplitudes. Indeed, the observed $m=-4$ wave is a factor of $\sim 2$ smaller in amplitude than our energy equipartition calculation predicts. If the amplitude of the $m=-5$ f-mode is smaller by a similar factor, it would not make an observable impact on the rings, in accordance with its non-observation.

\section{Discussion}
\label{discussion}

We have examined the oscillation mode spectrum of Saturn models with stable stratification deep in the interior. The stably stratified region supports g-modes, and we find that the g-mode spectrum extends to higher frequencies than the f-modes that generate observed waves in Saturn's rings. G-modes close in frequency to the f-modes mix with them, allowing them to obtain large enough gravitational perturbations to generate waves in the rings. We have proposed that this mechanism is responsible for the observed small frequency splittings between modes of the same azimuthal number $m$ inferred from ring observations. If true, this requires the existence of a thick region of stable stratification deep within the interior of Saturn, in contrast to the conventional notion that giant planets envelopes are convective throughout. 

Our models very naturally reproduce the observed frequency splitting of $\sim 5 \%$ between the two $m=-2$ waves observed by HN13. In fact, for g-mode cavities located deep within the planet near a ``core-envelope'' interface at $r/R \sim 0.25$, frequency splittings of this magnitude are difficult to avoid. Neither fully convective models nor models containing shear modes in a solid core can readily produce frequency splittings of this magnitude (see F14).

\subsection{Constraints on Saturn's Interior Structure}

Our results place some basic constraints on the interior structure of Saturn. In order to reproduce the measured frequencies of the f-modes, the core of Saturn $(r \lesssim 0.3 R)$ must be substantially more dense than Saturn's envelope. A dense core is also required by the measured vale of $J_2$, and it is reassuring that our seismic analysis is consistent with this constraint. More quantitative constraints are difficult because f-mode frequencies are only slightly affected by the properties of the core (F14). One must include rotational/ellipticity terms up to at least $(\Omega_s/\Omega_{\rm dyn})^4$ to calculate mode frequencies accurate enough to tightly constrain core properties.

We can also place some loose constraints on the characteristics of the stably stratified region within Saturn. In order for the g-modes to extend to high enough frequencies to mix with f-modes, the stable stratification must exist deep within the interior of Saturn. This is most naturally achieved for stable stratification at the large density gradient between the core and envelope at $r \sim 0.25 R$. We also find the stable stratification must be confined to relatively small radii $r \lesssim 0.5 R$. If the stable stratification extends to larger radii, mixing between $l=-m$ f-modes and g-modes seems to be too strong. In this case, we would expect the minimum frequency splitting between the $m=-2$ modes to be larger than the measured value of $\sim 5 \%$. We would also expect similar frequency splitting to be observed for both the $m=-3$ and $m=-4$ waves, in contrast with current observations.

The possible existence of stable stratification deep within giant planets has been speculated for decades (Stevenson 1985) and may be in accordance with recent theoretical investigations of giant planet interior structures. Wilson \& Militzer (2012a, 2012b) have claimed that ices and rocks will dissolve in liquid metallic hydrogen at temperatures expected for Saturn's core. This dissolution and ensuing diffusion of heavy elements could set up a stabilizing composition gradient near the core-envelope interface, similar to the stable stratification in our Saturn models. Leconte \& Chabrier (2012) have proposed that the counteracting effects of stabilizing composition gradients and destabilizing temperature gradients set up layered double-diffusive convective regions in giant planet interiors (see also Leconte \& Chabrier 2013). The stabilizing composition gradient leads to a real-valued Brunt-Vaisala frequency (at least when averaged over many layers) which could support g-modes. Alternatively, the layers may become unstable and merge (Rosenblum et al. 2011, Mirouh et al. 2012, Wood et al. 2013), producing much thicker regions of stability/convection. We have only investigated the effect of a single stably stratified region; the effect of alternating layers of convection/stratification on the g-mode spectrum is not clear.

\subsection{Explaining the Small Frequency Splittings}

Unfortunately, our models do not easily reproduce the frequency splittings of $\sim 0.3 \%$ between the three observed $m=-3$ waves. Although such splitting can occur, it is relatively rare, only occurring in roughly one tenth of our Saturn models. Below, we investigate several ideas that could explain this discrepancy.


{\it Possibility 1: the observed finely split waves are excited by a single mode.} There could be some dynamical mechanism that causes a single mode to excite multiple wave trains in the rings. This seems unlikely because splitting of this nature is not observed for waves excited at Lindblad resonances with Saturn's moons. Alternatively, the splitting could be produced if there is some mechanism that periodically modulates the f-mode amplitudes or phases on a timescale of $1/\delta \omega \sim $ weeks. There are no obvious dynamical/hydrodynamical mechanisms that could produce this sort of modulation.


{\it Possibility 2: our Saturn models are fundamentally different from Saturn's interior structure.} Perhaps our models lack a feature necessary to reproduce the observed mode spectrum. We have performed calculations with models containing a single density jump (which produces a single associated interface mode, for each value of $l$ and $m$) and have found no improvement over simpler models. However, it may be possible that models containing many density interfaces (or models including alternating convective/stratified layers produced by double-diffusive convection) could better match the observations. Alternatively, additional regions of stable stratification may exist within Saturn, e.g., at the molecular-metallic hydrogen phase transition or at a layer of helium accumulation due to helium rain out.


{\it Possibility 3: we have not included all the necessary pseudo-mode basis functions in our calculations.} It is possible that the inclusion of more basis functions in the matrix equation \ref{mat7} will lead to stronger mode mixing with the f-modes. We have performed various calculations that have included large $l$ modes $(l \gtrsim 30)$, high radial order g-modes, negative frequency (retrograde) modes, inertial modes, and Rossby modes. None of these additions appears to make a substantial qualitative difference in the mode spectrum near the prograde f-modes. Nonetheless, it may be possible that one must include a very large number of pseudo-modes to calculate an accurate normal mode spectrum. 

{\it Possibility 4: a different mechanism causes mode splitting.} The inclusion of magnetic fields, differential rotation, meridional circulation, non-spherical corrections to the planetary structure, non-adiabatic effects or some other bit of un-included physics could cause mode splitting. Magnetic effects are unlikely to be important (see discussion in Fuller et al. 2014). To the best of our knowledge, the other effects listed above can perturb mode frequencies but do not cause mode splitting.

{\it Possibility 5: mode mixing is enhanced by excluded physics.} The effects listed above may substantially increase the degree of mode mixing amongst pseudo-modes, allowing strong mixing between f-modes and high $l$ g-modes to be more common. Latitudinal differential rotation, in particular, will allow for direct mixing between pseudo-modes with $|l_\alpha - l_\beta| > 2$ and may make strong mixing more likely to occur. We find this option to be the most appealing out of all the possibilities listed in this section.

We anticipate that future studies can either eliminate the possibilities listed above or demonstrate their validity. The detection of additional mode-driven waves in the rings would also help narrow down the possibilities.

\subsection{Mode Amplitudes, Excitation Mechanisms, and Future Prospects}
\label{modeamp}

To determine mode amplitudes, we have simply normalized the mode amplitudes such that the $l=3$, $m=-3$ f-mode reproduces the largest observed $m=-3$ wave. This entails dimensionless mode amplitudes of $|A| \sim 2 \times 10^{-9}$, radial surface displacements of $U \sim 60$ cm, surface radial velocities of $v_r \sim 0.06$ cm/s, and mode energies of $E \sim 5 \times 10^{-18} GM^2/R$. These amplitudes are approximately the same as those calculated by MP93. The mechanism responsible for mode excitation remains unclear.

However, we note that energy equipartition amongst modes does not appear to be consistent with observations. Equipartition over-predicts the amplitude of the wave generated by the $l=4$, $m=-4$ f-mode, and predicts that the $l=5$, $m=-5$ f-mode should be observable (but it has not been detected). This implies that the excitation mechanism favors low frequency and/or low $l$ oscillation modes. The low frequency convective motions near Saturn's surface may be a good candidate for mode excitation.  

Although energy equipartition approximately predicts the correct amplitude of the largest $m=-2$ wave (Wave A in Table \ref{table}), it is possible that this wave is generated by a mixed g-mode slightly split in frequency from the $l=2$, $m=-2$ f-mode. The f-mode may instead create the Maxwell gap, as originally postulated by MP93 (see Figure \ref{m2}). This would require the f-mode be split in frequency by less than $1\%$ from the g-mode, similar to the splitting of the $m=-3$ modes. It would also imply the $m=-2$ modes have amplitudes larger than suggested by energy equipartition, in accordance with the arguments above.

Seismological constraints would be greatly improved with the detection of p-modes via radial velocity techniques. Unfortunately, if Saturn's p-modes have energies similar to those of the f-modes, the surface radial velocity variation for a p-mode with angular frequency $\sigma = 10$ mHz is only $v \sim 0.5$ cm/s. The apparent decrease of mode energy with frequency amongst Saturn's f-modes is not encouraging for this prospect. Nonetheless, we encourage searches for p-modes in Saturn and Jupiter, since they would provide valuable constraints on the interior structure (Jackiewicz et al. 2012).

We can also compare the mode amplitudes calculated here to those claimed to be observed in Jupiter via radial velocity techniques by Gaulme et al. (2011). To do this, we construct a very simple Jupiter model via the same method described in Section \ref{models}, which has a mass, radius, and $J_2$ that match those of Jupiter. We then calculate $l=2$ mode eigenfunctions for p-modes with frequencies $f_\alpha = \omega_\alpha/(2\pi) \approx 1.2$ mHz, similar to the frequencies of the Gaulme et al. (2011) modes. Finally, we calculate approximate mode amplitudes from the observed radial velocity variations (of order $v_{\rm ob} \sim 40$ cm/s) via $|A|_{\rm Jup} = v_{\rm ob}/[U_\alpha(R)\omega_\alpha]$.

This procedure results in p-mode amplitudes of $A_{\rm Jup} \sim 10^{-8}$ and energies of $E \sim 2 \times 10^{-14} GM_{\rm Jup}^2/R_{\rm Jup}$. As in Fuller et al. (2014), we find that the Gaulme et al. (2011) measurements entail modes that are more than a thousand times more energetic than the Saturn modes (compared to the gravitational binding energy of each planet). Such large p-mode amplitudes seem unlikely, especially in light of the fact that the higher frequency Saturn modes appear to have smaller energies. We therefore remain skeptical that the Gaulme et al. (2011) detections are truly due to Jupiter's p-modes. 

It may be possible for {\it Juno} to detect the gravitational influence of Jupiter's oscillation modes. The Gaulme et al. (2011) mode amplitudes imply potential perturbations of $\delta \Phi(R) \sim 10^{-10}$ (in units of Jupiter's gravitational potential) associated with Jupiter's p-modes. However, if Jupiter's f-modes have similar energies, they will generate potential perturbations of $\delta \Phi(R) \sim 10^{-7}$ at periods of $\sim$ hours. {\it Juno} may be able to detect these gravitational anomalies (Kaspi et al. 2010). If, however, Jupiter's f-modes have similar energies to those of Saturn (in units of the planetary binding energy), the modes only produce potential perturbations of $\delta \Phi(R) \sim 2 \times 10^{-9}$.

\subsection{Predictions}

Our claim that Saturn contains stable stratification deep within its interior makes several testable predictions. First, our models predict that the observed modes are only the ``tip of the ice berg", and that Saturn hosts a dense spectrum of g-modes in the f-mode frequency regime (see Figures \ref{m2}-\ref{m4}) that are currently unobserved. Observing these modes through their effect on the rings may be impossible because their gravitational perturbations are too weak to launch detectable waves in the rings. Their surface displacements are correspondingly small because the modes are localized in the g-mode cavity deep within the planet, so observing them with radial velocity measurements may also be very difficult. Nonetheless, their detection would be consistent with our theory.

Second, our models predict that there may be other, small amplitude waves in the rings excited by $l=-m$ g-modes. For instance, Figure \ref{m2} shows an $m=-2$ g-mode with a Lindblad resonance at $\sim 9.3 \times 10^{4}$km that is on the border of detectability. The exact frequencies of the detectable g-modes are not robust predictions of our model, we merely speculate that there could be other low amplitude $|m|\sim2-4$ waves separated from the observed waves by a few percent in frequency. There are several unidentified waves in the C-ring (see Marley 2014) that could be excited by these g-modes. 

Finally, we find the arguments presented in the preceding subsection compelling enough to predict that the Maxwell gap is generated by the $l=2$, $m=-2$ f-mode of Saturn, as originally proposed in MP93. Indeed, there is an observed wave train on the eccentric ringlet within the Maxwell gap (Porco et al. 2005, Nicholson et al. 2014, in prep), which may be excited by the $l=2$, $m=-2$ f-mode. We predict this wave train has $m=-2$ and that its pattern frequency is $\Omega_p \simeq 1770 ^\circ/{\rm day}$.

\section*{Acknowledgments} 

I thank Matt Hedman, Phil Nicholson, Yanqin Wu, Dave Stevenson, Mark Marley, and Gordon Ogilvie for helpful discussions. I acknowledge partial support from NSF under grant no. AST-1205732 and through a Lee DuBridge Fellowship at Caltech. This research was supported in part by the National Science Foundation under Grant No. NSF PHY11-25915.

\appendix

\section{Ellipticity}
\label{ellipticity}

Here we summarize our method for calculating the effect of rotation on the structure of our Saturn models. We adopt a perturbative method, considering only terms of order $(\Omega_s/\Omega_{\rm dyn})^2$. Our method follows Eggelton 2006. In the perturbative method, the radius of an elliptical shell is
\beq
\label{rac2}
r_{\rm ac} = r \big[ 1 - \varepsilon(r) P_2 (\cos \theta) \big],
\eeq
where $r$ is the corresponding radius in the spherical model, and $P_2$ is the $l=2$ Legendre polynomial. The relative ellipticity $\epsilon(r) = \varepsilon(r)/\varepsilon(R)$ is found by solving Clairaut's equation:
\beq
\label{clairaut}
\frac{d^2}{dr^2} \epsilon + \frac{8\pi G \rho}{g} \bigg( \frac{d}{dr} + \frac{1}{r} \bigg) \epsilon - \frac{6}{r^2}\epsilon = 0.
\eeq
The central boundary condition is $d\epsilon/dr = 0$ at $r=0$, while the definition of $\epsilon$ requires $\epsilon=1$ at $r=R$. 

The value of $\varepsilon$ must be normalized according to the value of $(\Omega_s/\Omega_{\rm dyn})^2$. To do this, we first compute the integral
\beq
\label{Qclair}
Q = \frac{1}{5M R^2} \int^R_0 8 \pi \rho r^4 \big(5 \epsilon + r d \epsilon/dr \big)dr.
\eeq
Then the surface ellipticity is
\beq
\varepsilon(R) = \frac{\Omega_s^2}{3 \Omega_{\rm dyn}^2} \frac{1}{1-Q}
\eeq
and the gravitational moment $J_2$ is
\beq
J_2 = \frac{\Omega_s^2}{3 \Omega_{\rm dyn}^2} \frac{Q}{1-Q}.
\eeq
Note that both the ellipticity $\varepsilon(r)$ and the value of $J_2$ are proportional to the small number $(\Omega_s/\Omega_{\rm dyn})^2$.

\section{Solving for pseudo-modes}
\label{pseudo}

Here we describe our method to solve for the pseudo-modes of the planet, which will be used as the basis functions for our final mode calculation. Our method uses similar techniques to the analysis of Ogilvie \& Lin (2004). We begin by examining the linearized adiabatic fluid momentum equation
\beq
\label{momeq}
-\rho \omega^2 \bxi = - \bnab \delta P  - \rho \bnab \delta \Phi  - g \delta \rho {\hat {\bf r}} - 2 i \rho \omega {\bf \Omega}_s \times \bxi.
\eeq
for a perturbation with time dependence 
\beq
\label{conv}
\bxi \propto e^{i(\omega t + m \phi)}.
\eeq
Equation \ref{momeq} (and all subsequent analysis) applies in the rotating frame. The time dependence of equation \ref{conv} implies the prograde modes we will be interested in have $m<0$. The perturbation variables $\bxi$, $\delta \rho$, $\delta P$, and $\delta \Phi$ are the Lagrangian displacement and Eulerian perturbations in density, pressure and gravitational potential, respectively. We consider solid body rotation such that the angular spin frequency is ${\bf \Omega}_s = \Omega_s \hat{{\bf z}}$. Equation \ref{momeq} applies to a spherical planetary model, we will introduce centrifugal/oblateness effects in a perturbative manner in Section \ref{matrix}. We also use the continuity equation 
\beq
\label{cont}
\delta \rho + \frac{1}{r^2} \frac{\partial}{\partial r} \big( \rho r^2 \xi_r \big) + \rho \bnab_\perp \cdot \bxi_\perp = 0,
\eeq
the adiabatic equation of state
\beq
\label{drho}
\delta \rho = \frac{1}{c_s^2} \delta P + \frac{\rho N^2}{g} \xi_r,
\eeq
and Poisson's equation
\beq
\label{poisson}
\frac{1}{r^2}\frac{\partial}{\partial r} \bigg( r^2 \frac{\partial}{\partial r} \delta \Phi \bigg) + \nabla_\perp^2 \delta \Phi = 4 \pi G \delta \rho.
\eeq
Here, $c_s$ is the adiabatic sound speed, $N$ is the Brunt-Vaisala frequency, and G is Newton's gravitational constant. 

We now project the pseudo-modes onto spherical harmonics. We choose
\beq
\label{xir}
\xi_r (r, \theta, \phi) = U(r) Y_{lm}(\theta, \phi),
\eeq
\beq
\label{xit}
\xi_\theta (r, \theta, \phi) = V(r) \frac{\partial}{\partial \theta} Y_{lm}(\theta, \phi) + i W(r) \frac{1}{\sin \theta} \frac{\partial}{\partial \phi} Y_{l+1, m}(\theta, \phi)
\eeq
\beq
\label{xip}
\xi_\phi (r, \theta, \phi) = V(r) \frac{1}{\sin \theta} \frac{\partial}{\partial \phi} Y_{lm}(\theta, \phi) - i W(r) \frac{\partial}{\partial \theta} Y_{l+1, m}(\theta, \phi),
\eeq
\beq
\label{psi}
\frac{\delta P (r, \theta, \phi)}{\rho} + \delta \Phi(r, \theta, \phi) = \Psi(r) Y_{lm}(\theta, \phi),
\eeq
\beq
\label{phi}
\delta \Phi(r, \theta, \phi) = \delta \Phi(r) Y_{lm}(\theta, \phi).
\eeq
In equations \ref{xit} and \ref{xip}, the horizontal displacements $V(r)$ and $W(r)$ represent the poloidal and toroidal parts of the horizontal displacement, respectively. We have chosen the toroidal piece to have $Y_{l+1,m}$ dependence so that it has the same symmetry as the poloidal component, therefore a pseudo-mode with an even value of $l+m$ is symmetric across the equator, whereas a pseudo-mode with an odd value of $l+m$ is anti-symmetric across the equator. Also note that our definition of the toroidal displacement $W$ is different by a factor of $i$ from Ogilvie \& Lin 2004 and DT98. Our choice insures that the values of $U$, $V$, $W$, $\Psi$, and $\delta \Phi$ are purely real.

As we shall see below, the pseudo-modes obtained from the projections \ref{xir}-\ref{phi} are not actual solutions of the momentum equation \ref{momeq}. They will satisfy equation \ref{momeq} only after an integration over angle that eliminates coupling between spherical harmonics of different $l$. This coupling will be restored when we account for mixing between pseudo-modes of different $l$ in \ref{matrix}.

Equation \ref{momeq} is, dropping the coordinate dependence of the variables for convenience
\begin{align}
\label{U}
&- \rho \omega^2 U Y_{lm} = -\frac{\partial}{\partial r} \Psi Y_{lm} + \frac{N^2}{g} \left( \Psi - \delta \phi \right) Y_{lm} - \rho N^2 U Y_{lm} \nonumber \\
&+ i q \rho \omega^2 \left( V \frac{\partial}{\partial \phi} Y_{lm} - i W \sin \theta \frac{\partial}{\partial \theta} Y_{l+1,m} \right),
\end{align}
\begin{align}
\label{V}
&- \rho \omega^2 \left( V \sin \theta \frac{\partial}{\partial \theta} Y_{lm} + i W \frac{\partial}{\partial \phi} Y_{l+1,m} \right) = \frac{-1}{r} \Psi \sin \theta \frac{\partial}{\partial \theta} Y_{lm} \nonumber \\
&+ i q \rho \omega^2 \left( V \cos \theta \frac{\partial}{\partial \phi} Y_{lm} - i W \sin \theta \cos \theta \frac{\partial}{\partial \theta} Y_{l+1,m} \right),
\end{align}
\begin{align}
\label{W}
&- \rho \omega^2 \left( V \frac{\partial}{\partial \phi} Y_{lm} - i W \sin \theta \frac{\partial}{\partial \theta} Y_{l+1,m} \right) = \frac{-1}{r} \Psi  \frac{\partial}{\partial \phi} Y_{lm} \nonumber \\
&- i q \rho \omega^2 \left( V \sin \theta \cos \theta \frac{\partial}{\partial \theta} Y_{lm} + i W \cos \theta \frac{\partial}{\partial \phi} Y_{l+1,m}  + U \sin^2 \theta Y_{lm} \right),
\end{align}
where
\beq
\label{q}
q = \frac{2 \Omega_s}{\omega}
\eeq
is the rotation parameter. Additionally, the continuity equation is
\beq
\label{cont2}
\frac{\rho}{c_s^2} \left(\Psi - \delta \Phi \right) Y_{lm} + \frac{\rho N^2}{g} U Y_{lm} + \frac{1}{r^2} \frac{\partial}{\partial r} \left( \rho r^2 U \right) Y_{lm} - \frac{l(l+1) \rho}{r} V Y_{lm} = 0.
\eeq
To simplify equations \ref{V} and \ref{W}, we exploit the fact that the divergence of a toroidal field is zero, while the curl of a poloidal field is zero. To do this, we operate by $(1/\sin \theta) \partial /\partial \theta$ on equation \ref{V} and add it to $(i/\sin^2 \theta) \partial /\partial \phi$ operated on equation \ref{W}. Using $\nabla^2 Y_{lm} = (-l(l+1)/r^2) Y_{lm}$ and $\partial Y_{lm}/\partial \phi = i m Y_{lm}$, we have
\begin{align}
\label{V1}
l(l+1) V Y_{lm} &= \frac{l(l+1)}{r \omega^2} \Psi Y_{lm} + m q V Y_{lm} + m q U Y_{lm} \nonumber \\
&- q W \left[ (l+1)(l+2) \cos \theta Y_{l+1, m} + \sin \theta \frac{\partial}{\partial \theta} Y_{l+1,m} \right].
\end{align}
Then, we operate on equation \ref{V} by $(i/\sin^2 \theta) \partial/\partial \phi $ and combine it with $(1/\sin \theta) \partial /\partial \theta$ operated on equation \ref{W} to find
\begin{align}
\label{W1}
(l+1)(l+2) W Y_{l+1,m} &= m q W Y_{l+1,m} - q V \left[ l(l+1) \cos \theta Y_{lm} + \sin \theta \frac{\partial}{\partial \theta} Y_{lm} \right] \nonumber \\ &+ q U \left[ 2 \cos \theta Y_{lm} + \sin \theta \frac{\partial}{\partial \theta} Y_{lm} \right] .
\end{align}

We now use the identities
\beq
\label{ylm1}
\cos \theta Y_{lm} = S_{lm} Y_{l-1, m} + S_{l+1, m} Y_{l+1, m},
\eeq
with
\beq
\label{slm}
S_{lm} = \left[ \frac{ (l+m)(l-m)}{(2l+1)(2l-1)} \right],
\eeq
and
\beq
\label{ylm2}
\sin \theta \frac{\partial}{\partial \theta} Y_{lm} = l S_{l+1,m} Y_{l+1, m} - (l+1) S_{l m} Y_{l-1, m}.
\eeq
Then, multiplying equation \ref{V1} by $Y_{lm}^*$ and equation \ref{W1} by $Y_{l+1,m}^*$ and integrating over angle, we find
\beq
\label{V2}
l(l+1) V  - m q V = \frac{l(l+1)}{r \omega^2} \Psi  + m q U - q l(l+2)S_{l+1,m} W,
\eeq
and
\beq
\label{W2}
(l+1)(l+2) W - m q W =  q (l+2)S_{l+1,m} U - q l(l+2)S_{l+1,m} V 
\eeq
The angular integration has eliminated terms proportional to $Y_{l\pm2,m}$ that cause mixing between pseudo-modes of different $l$, and which will be accounted for in \ref{matrix}. Equations \ref{V2} and \ref{W2} provide algebraic relations for the values of $V$ and $W$ at any radius $r$. Inspection of equations \ref{V2} and \ref{W2} reveal that in the non-rotating limit $q \rightarrow 0$ we obtain $W=0$, i.e., the displacements are purely poloidal. Furthermore, when $q \rightarrow 0$, we recover $V = [\Psi/(r \omega^2)] $ which is the algebraic relation typically used in computations of modes in non-rotating bodies.

We can perform similar angular integrations on equations \ref{cont}, \ref{poisson}, and \ref{U}. We are left with a system of six equations, stemming from the three momentum equations, the second order Poisson's equation, and the continuity equation. These equations are composed of four differential equations and two algebraic relations for the six unknowns $U$, $V$, $W$, $\Psi$, $\delta \Phi$, and $\delta g$:
\begin{align}
\label{eqn1}
& \frac{\partial}{\partial r} \Psi - \frac{N^2}{g} \big( \Psi - \delta \Phi \big) + \big( N^2 - \omega^2 \big) U + q m \omega^2 V + q(l+2) S_{l+1, m} \omega^2 W = 0 \\
\label{eqn2}
& \frac{\partial}{\partial r} U + \left( \frac{2}{r} - \frac{g}{c_s^2} \right) U + \frac{1}{c_s^2} \big( \Psi - \delta \Phi \big) - \frac{l(l+1)}{r} V = 0 \\
\label{eqn3}
& \frac{\partial}{\partial r} \delta \Phi - \delta g = 0 \\
\label{eqn4}
& \frac{\partial}{\partial r} \delta g + \frac{2}{r} \delta g - \frac{l(l+1)}{r^2} \delta \Phi - 4 \pi G \rho \left[ \frac{1}{c_s^2} \big( \Psi - \delta \Phi \big) + \frac{N^2}{g} U \right] = 0\\
\label{eqn5}
& \left[ l(l+1) - m q \right] V - \frac{l(l+1)}{r \omega^2} \Psi - m q U + q l(l+2) S_{l+1, m} W = 0 \\
\label{eqn6}
& \left[ (l+1)(l+2) - m q \right] W + ql(l+2)S_{l+1,m} V - q (l+2) S_{l+1,m} U = 0
\end{align} 
Of course, equations \ref{eqn1}-\ref{eqn6} also contain the additional unknown eigenfrequency $\omega$, which satisfies a trivial seventh equation 
\beq
\label{omeqn}
\frac{\partial}{\partial r} \omega = 0.
\eeq

In order to solve the seven equations \ref{eqn1}-\ref{omeqn}, we require seven boundary conditions. For pseudo-modes with $l\geq 2$, the usual boundary conditions apply:
\begin{align}
\label{inbc1}
& U - l V = 0 \ \ {\rm at} \ r \rightarrow 0 \\
\label{inbc2}
&  r \delta g - l \delta \Phi = 0 \ \ {\rm at} \ r \rightarrow 0 \\
\end{align}
and 
\begin{align}
\label{outbc1}
& \Psi - \delta \Phi - g U = 0 \ \ {\rm at} \ r \rightarrow R \\
\label{outbc2}
&  r \delta g + (l+1) \delta \Phi = 0 \ \ {\rm at} \ r \rightarrow R \\.
\end{align}
The algebraic relations \ref{eqn5} and \ref{eqn6} constitute two more boundary conditions when applied at $r \rightarrow 0$ or $r \rightarrow R$.

The final boundary condition is a normalization condition. The most common choice is $U=1$ at $r=R$. However, the normalization is entirely arbitrary, and we use different normalizations to find different types of modes. One must be cautious with this choice, as the modes have very different scales (the g-modes, for instance, are localized in the g-mode cavity and have very small perturbations near the surface, whereas the opposite is true of inertial modes), and it is difficult to choose a single normalization that allows our relaxation code to reliably find all classes of modes.

After we solve for the pseudo-modes, we renormalize them using the condition
\beq
\label{normpseudo}
\int dV \rho \bxi^* \cdot \bxi - \frac{i}{\omega} \int dV  \rho \bxi^* \cdot \big( {\boldsymbol \Omega}_s \times \bxi \big) = 1
\eeq
The reason for this choice is that the pseudo-modes are orthonormal to one another under equation \ref{normpseudo} for a given $l$ (see \ref{matrix}). However, pseudo-modes of different $l$ are coupled to one another by the Coriolis force, and we must use the technique described in  \ref{matrix} to solve for the normal modes.

\section{Mode Eigensystem}
\label{matrix}

Here we describe our method for solving the mode eigensystem to determine mode eigenfrequencies $\omega$ and eigenfunctions $\bxi$, using the pseudo-modes as basis functions. We follow the procedure outlined in Dahlen \& Tromp 1998 (DT98). An oscillation mode is a solution to the generalized eigenvalue problem
\beq
\label{mat1}
\begin{bmatrix} 0 & \mathcal{V} \\ \mathcal{V} & 2 \mathcal{W} \end{bmatrix} {\bf z} = \omega \begin{bmatrix} \mathcal{V} & 0 \\ 0 & \mathcal{T} \end{bmatrix} {\bf z},
\eeq
where 
\beq
\label{z}
{\bf z} = \left[ \begin{array}{c} {\bxi} \\ \omega {\bxi} \end{array} \right].
\eeq
the $\mathcal{V}$, $\mathcal{T}$, $\mathcal{W}$ operators correspond to potential energy, kinetic energy, and Coriolis force operators. These operators are most conveniently expressed through their inner products with the pseudo-mode displacements:
\begin{align}
\label{tmat}
T_{\alpha \beta} &= \langle \bxi_\alpha | \mathcal{T} | \bxi_\beta \rangle \nonumber \\
&= \int dV \rho \bxi_\alpha^* \cdot \bxi_\beta,
\end{align}
\begin{align}
\label{wmat}
W_{\alpha \beta} &= \langle \bxi_\alpha | \mathcal{W} | \bxi_\beta \rangle \nonumber \\
&= \int dV \rho \bxi_\alpha^* \cdot \big( i {\boldsymbol \Omega}_s \times \bxi_\beta \big)
\end{align}
\begin{align}
\label{vmat}
V_{\alpha \beta} &= \langle \bxi_\alpha | \mathcal{V} | \bxi_\beta \rangle,
\end{align}
with $V_{\alpha \beta}$ given by equation 7.36 of DT98. With our choice of the definition of $W$ (equations \ref{xit} and \ref{xip}), all three operators are real and symmetric such that $T_{\alpha \beta} = T_{\beta \alpha}$ and likewise for $V_{\alpha \beta}$ and $W_{\alpha \beta}$. 

In the non-rotating limit, equation \ref{mat1} reduces to the more familiar eigensystem, $\mathcal{V} \bxi = \omega^2 \mathcal{T} \bxi$ with the orthonomality requirement $T_{\alpha \beta} = \delta_{\alpha \beta}$. Including rotation, the modes must satisfy $\big[ \mathcal{V} + 2 \omega \mathcal{W} \big] \bxi= \omega^2 \mathcal{T} \bxi$ with the modified orthonormality relation 
\beq
\label{orthonorm}
T_{\alpha \beta} = \delta_{\alpha \beta} + \frac{2}{\omega_\alpha + \omega_\beta} W_{\alpha \beta}. 
\eeq

The pseudo-modes we calculate in \ref{pseudo} are not normal modes because they do not satisfy equation \ref{mat1}, nor do they satisfy the orthonormality relation \ref{orthonorm} for $l_\alpha \neq l_\beta$. However, the pseudo-modes do satisfy the relation 
\beq
\label{pseudonorm}
\langle \bxi_\alpha | \mathcal{V} + 2 \omega_\alpha \mathcal{W} - \omega_\alpha^2 \mathcal{T} | \bxi_\alpha \rangle = 0.
\eeq
Furthermore, for $l_\alpha = l_\beta$, the pseudo-modes are orthogonal in the sense that they satisfy 
\beq
\label{pseudoorth}
\langle \bxi_\alpha | \mathcal{V} + 2 \omega_\beta \mathcal{W} - \omega_\beta^2 \mathcal{T} | \bxi_\beta \rangle  - \langle \bxi_\beta | \mathcal{V} + 2 \omega_\alpha \mathcal{W} - \omega_\alpha^2 \mathcal{T} | \bxi_\alpha \rangle = 0 \ \ {\rm for} \ \alpha \neq \beta.
\eeq
Equations \ref{pseudonorm} and \ref{pseudoorth} can be verified with a fair amount of algebra from equations \ref{U}-\ref{W} and \ref{eqn1}-\ref{eqn6}. The pseudo-modes therefore satisfy, using our chosen normalization of equation \ref{normpseudo},
\begin{align}
\label{tpseudo}
T_{\alpha \beta} &= \delta_{\alpha \beta} + \frac{2}{\omega_\alpha + \omega_\beta} W_{\alpha \beta} \ \ {\rm for} \ l_\alpha = l_\beta \nonumber \\
&= 0 \ \ {\rm for} \ l_\alpha \neq l_\beta 
\end{align}
and 
\begin{align}
\label{vpseudo}
V_{\alpha \beta} &= \omega_\alpha^2 \delta_{\alpha \beta} - \frac{2\omega_\alpha \omega_\beta}{\omega_\alpha + \omega_\beta} W_{\alpha \beta} \ \ {\rm for} \ l_\alpha = l_\beta \nonumber \\
&= 0 \ \ {\rm for} \ l_\alpha \neq l_\beta.
\end{align}
The second lines of equations \ref{tpseudo} and \ref{vpseudo} can be easily verified from equations \ref{tmat} and \ref{vmat} because the operators $\mathcal{T}$ and $\mathcal{V}$ do not induce coupling between spherical harmonics of different values of $l$. However, the pseudo-modes are coupled across different values of $l$ because $W_{\alpha \beta} \neq 0$ for $l_\alpha \neq l_\beta$. 

Introducing centrifugal/ellipticity effects modifies equation \ref{mat1} to 
\beq
\label{matpert}
\begin{bmatrix} 0 & \mathcal{V} + \delta \mathcal{V} \\ \mathcal{V} + \delta \mathcal{V}& 2 \mathcal{W} \end{bmatrix} {\bf z} = \omega \begin{bmatrix} \mathcal{V} + \delta \mathcal{V}& 0 \\ 0 & \mathcal{T} + \delta \mathcal{T} \end{bmatrix} {\bf z}.
\eeq
The form of the inner products $\delta V_{\alpha \beta}$ and $\delta T_{\alpha \beta}$ are quite lengthy. We refer the reader to equations D.80-D.97 of DT98 for explicit formulae. However, it is important to note some differences in notation. The relationships between the variables in DT98 and our variables are $\varepsilon_{\rm DT} = (3/2) \varepsilon$, and $W_{\rm DT} = -i W$. In short, these terms introduce mixing between a pseudo-mode of $l_\alpha$ with other pseudo-modes of $l_\beta = l_\alpha$ and $l_\beta = l_\alpha \pm 2$. Since the axial symmetry is maintained, modes only mix if $m_\alpha = m_\beta$.

To solve equation \ref{matpert} for $\omega$, ${\bf z}$ eigenvalue combinations, we decompose ${\bf z}$ into our pseudo-mode basis:
\beq
\label{xidecomp}
{\bf z} = \sum_\beta a_\beta {\bf z}_\beta.
\eeq
Inserting the expansion \ref{xidecomp} into equation \ref{matpert}, we obtain
\beq
\label{mat2}
\sum_\beta a_\beta \left( \begin{bmatrix} 0 & \mathcal{V} + \delta \mathcal{V} \\ \mathcal{V} + \delta \mathcal{V}& 2 \mathcal{W} \end{bmatrix} \right ) \left[ \begin{array}{c} {\bxi}_\beta \\ \omega_\beta {\bxi}_\beta \end{array} \right] = \omega \sum_\beta a_\beta \begin{bmatrix} \mathcal{V} + \delta \mathcal{V}& 0 \\ 0 & \mathcal{T} + \delta \mathcal{T} \end{bmatrix} \left[ \begin{array}{c} {\bxi}_\beta \\ \omega_\beta {\bxi}_\beta \end{array} \right].
\eeq
We now take the inner product of equation \ref{mat2} with a basis mode conjugate ${\bf z}_\alpha^H = {{\bf z}_\alpha^*}^{T}$ to find
\begin{align}
\label{mat3}
&\sum_\beta a_\beta \left[ (\omega_\alpha + \omega_\beta ) V_{\alpha \beta} + (\omega_\alpha + \omega_\beta) \delta V_{\alpha \beta} + 2 \omega_\alpha \omega_\beta W_{\alpha \beta} \right] \nonumber \\
&= \omega \sum_\beta a_\beta \left[ V_{\alpha \beta} + \delta V_{\alpha \beta} + \omega_\alpha \omega_\beta T_{\alpha \beta} + \omega_\alpha \omega_\beta \delta T_{\alpha \beta} \right] .
\end{align}
Using the pseudo-mode overlap equations \ref{tpseudo} and \ref{vpseudo}, we have 
\begin{align}
\label{mat4}
&2 \omega_\alpha^3 a_\alpha + \sum_\beta a_\beta \left[ 2 \omega_\alpha \omega_\beta W_{\alpha \beta} \big(1 - \delta_{l_\alpha l_\beta} \big) + (\omega_\alpha + \omega_\beta) \delta V_{\alpha \beta} \right] \nonumber \\
&= \omega \left( 2 \omega_\alpha^2 a_\alpha + \sum_\beta a_\beta \left[ \delta V_{\alpha \beta} + \omega_\alpha \omega_\beta \delta T_{\alpha \beta} \right] \right).
\end{align}
Letting $b_\beta = \omega_\beta a_\beta$, and dividing by $2 \omega_\alpha$, we have 
\beq
\label{mat5}
\omega_\alpha b_\alpha + \sum_\beta b_\beta \left[ W_{\alpha \beta} \big(1 - \delta_{l_\alpha l_\beta} \big) + \frac{\omega_\alpha + \omega_\beta}{2\omega_\alpha \omega_\beta} \delta V_{\alpha \beta} \right] = \omega \left( b_\alpha + \sum_\beta b_\beta \left[ \frac{1}{2\omega_\alpha \omega_\beta} \delta V_{\alpha \beta} + \frac{1}{2} \delta T_{\alpha \beta} \right] \right).
\eeq
Equation \ref{mat5} is a generalized Hermitian eigenvalue problem of form 
\beq
\label{mat6}
{\rm A} {\bf b} = \omega {\rm B} {\bf b},
\eeq
with the matrix elements of ${\rm A}$ given by
\beq
{\rm A}_{\alpha \beta} = \omega_\alpha \delta_{\alpha \beta} + \big( 1 - \delta_{l_\alpha l_\beta} \big) W_{\alpha \beta} + \frac{\omega_\alpha + \omega_\beta}{2 \omega_\alpha \omega_\beta} \delta V_{\alpha \beta}
\eeq
and ${\rm B}$ given by
\beq
{\rm B}_{\alpha \beta} = \delta_{\alpha \beta} + \frac{1}{2 \omega_\alpha \omega_\beta} \delta V_{\alpha \beta} + \frac{1}{2} \delta T_{\alpha \beta}. 
\eeq
The diagonal elements $\omega_\alpha$ are the unperturbed pseudo-mode eigenfrequencies. The matrices $W_{\alpha \beta}$, $\delta V_{\alpha \beta}$, and $\delta T_{\alpha \beta}$ induce mixing between pseudo-modes of $l_\alpha = l_\beta$ and $l_\alpha = \l_\beta \pm 2$. 

After calculating the matrices ${\rm A}$ and ${\rm B}$, we numerically solve the matrix equation
\beq
\label{mat7}
{\rm B}^{-1} {\rm A} {\bf b} = \omega {\bf b},
\eeq
for the eigenvalues $\omega$ and eigenvectors ${\bf b}$. This method is prone to numerical inaccuracies because the matrix $B^{-1} A$ is not symmetric. However, as long as the matrix $B$ is positive-definite (which it must be for secularly stable planetary models), one can use a Cholesky decomposition to solve the matrix equation \ref{mat6} (see Press et al. 1998). This decomposition ensures that the matrix involved is symmetric (making it more amenable for numeric solving algorithms) and hence that the eigenvalues are all real. In practice, we use both the Cholesky method and equation \ref{mat7}, and check that the results are identical. 

Upon solving equation \ref{mat6}, we normalize each mode via
\beq
\label{normmode}
\langle {\bf z} \vert \mathcal{P} \vert {\bf z} \rangle = 2 \omega^2,
\eeq
with 
\beq
\label{normP}
\mathcal{P} = \begin{bmatrix} \mathcal{V} + \delta \mathcal{V}& 0 \\ 0 & \mathcal{T} + \delta \mathcal{T} \end{bmatrix}.
\eeq

We caution that the inclusion of very low frequency modes (such as the Rossby modes) leads to numerical instability. The reason is that the value of $\delta V_{\alpha \beta}/(\omega_\alpha \omega_\beta)$ can become large for low frequency modes if the value of $\delta V_{\alpha \beta}$ has numerical error. This is nearly unavoidable on a finite radial grid since the value of $\delta V_{\alpha \beta}$ is computed from radial integrals over oscillatory mode eigenfunctions. Although this issue does not appear to affect our calculations at higher frequencies, our techniques may be numerically ill-suited for reliable calculations of very low frequency modes.

\end{document}